\documentclass[twocolumn,superscriptaddress,showpacs,pra,a4paper]{revtex4}
\usepackage{amsmath,amsfonts,bm,ifpdf}
\usepackage[pdftex]{graphicx}
\graphicspath{{figures/pdf/}}
\DeclareGraphicsExtensions{.pdf}

\def\op#1{\hat{#1}}
\def\vec#1{\mathbf{#1}}
\def\gvec#1{{\bm #1}}
\def\op#1{#1}
\def\ket#1{| #1 \rangle}
\def\bra#1{\langle #1 |}
\def\ip#1#2{\langle #1| #2 \rangle}
\def\ave#1{\langle #1 \rangle}
\def\norm#1{|| #1 ||}

\def\ONE{\mathbb{I}}

\def\U{\mathbb{U}}
\def\E{\mathcal{E}}
\def\diag{\mathop{\rm diag}}
\def\angle{\mathop{\rm phase}}
\def\opt{\rm opt}
\def\act{\rm act}
\def\est{\rm est}
\def\eps{\epsilon}

\newif\ifpdflatex\pdflatextrue
\makeatletter\@ifundefined{pdfoutput}{\pdflatexfalse}\makeatother
\def\myincludegraphics[#1]#2#3{%
\ifpdflatex \includegraphics[#1]{#2}
\else       \includegraphics[#1]{#3}
\fi}

\begin{document}

\bibliographystyle{prsty}

\title{Two-Qubit Hamiltonian Tomography by Bayesian
Analysis of Noisy Data}

\author{Sonia G.~Schirmer} \email{sgs29@cam.ac.uk}
\affiliation{Department of Applied Mathematics and Theoretical Physics,
             University of Cambridge, Wilberforce Road, Cambridge, CB3 0WA, UK}
\affiliation{Centre for Quantum Technologies, National University of Singapore,
             3 Science Drive 2, Singapore 117543}

\author{Daniel K.~L.~Oi}
\affiliation{SUPA, Department of Physics, University of Strathclyde, Glasgow G4
  0NG, UK} 
\affiliation{Centre for Quantum Technologies, National University of Singapore,
             3 Science Drive 2, Singapore 117543}
\date{\today}

\begin{abstract}
  We present an empirical strategy to determine the Hamiltonian dynamics of a
  two-qubit system using only initialization and measurement in a single fixed
  basis. Signal parameters are estimated from measurement data using Bayesian
  methods from which the underlying Hamiltonian is reconstructed, up to three
  unobservable phase factors.  We extend the method to achieve full control
  Hamiltonian tomography for controllable systems
  via a multi-step approach.  The technique is demonstrated and evaluated by
  analyzing data from simulated experiments including projection noise.
\end{abstract}


\maketitle

\section{Introduction}

\begin{figure}
\includegraphics[width=\columnwidth]{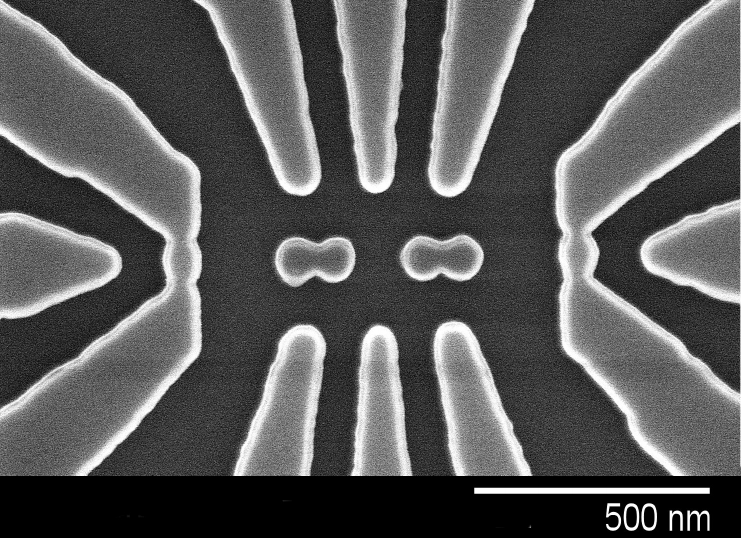}
\caption{Manufactured Qubit System. A pair of horizontally aligned double
  quantum dots (center) can act as a two-qubits. A qubit can be defined in each
  double quantum dot by two different charging states, e.g. a single excess
  electron located on the left or right dot of each pair. Electrodes (top and
  bottom) control the potentials and electron tunneling rates. Single electron
  transistors (left and right) measure the locations of the excess electrons
  which defines the measurement basis, or logical states of the qubit. Due to
  finite manufacturing precision, the placement of the control and measurement
  structures may not be exactly as calculated, hence the Hamiltonian dependence
  upon control signals will have to be determined empirically. Image courtesy of
  Hitachi Cambridge Laboratories, Hitachi Europe Ltd.}
\label{fig:SiQubits}
\end{figure}

Using quantum phenomena to perform new modes of computation is a daunting
challenge~\cite{qcomp}. Significant achievements in the theory of quantum
computation include the development of error correction,
fault-tolerance~\cite{error}, and scalability of quantum
circuits~\cite{scalability}. However, in order to build large scale quantum
processors, many individual quantum systems must be manipulated with
extraordinary precision and accuracy.  A prerequisite for this level of quantum
control is precise characterization of the underlying dynamics and its response
to control fields, so-called Hamiltonian Engineering~(\cite{hamileng1,hamileng2}
and references therein).  This is especially crucial for manufactured devices
such as solid state quantum bits (qubits), e.g. quantum dots
(Fig.~\ref{fig:SiQubits}) or superconducting quantum interference devices
(SQUIDs).  Any manufacturing process will introduce variations so it is
important to empirically identify the control relationship for each
component. In a large-scale quantum computer, it is desirable to be able to
achieve this using \textit{in situ} resources, i.e., initialization, control
actuators and measurement capabilities already present for performing
computation.

The canonical method for assessing quantum dynamics is quantum process
tomography (QPT)~\cite{qpt1,qpt2,qpt3}.  This involves initialization of
a quantum system in a (complete) set of states, allowing it to evolve
under the dynamics under consideration, and then performing an
informationally complete measurement on the output state for each
input. From this set of input-output data, the superoperator, or
completely positive (CP) map, governing the quantum evolution of the
system can be reconstructed. This may then be repeated for different
evolution times to obtain an estimate of the Lindblad operators
(generators of the dynamics)~\cite{lindblad}. For control purposes, QPT
would be performed for a variety of actuator settings to build up a map
of the control space.

A potential disadvantage of QPT is the need for \textit{ab initio}
initialization and measurement outside of the computational basis, a
capability which may not exist in the absence of characterization in the
first place. It is usually argued that initialization and measurement in
an arbitrary basis can be achieved by unitary rotation of a fixed basis,
however this pre-supposes that the system response to control fields has
already been characterized, a vicious circle. Previous work has
addressed this issue for the case of a single qubit subject to multiple
control Hamiltonians, decoherence, and imperfect subspace
confinement~\cite{SKOC2004,SKO2004,CSGWOH2005,CGOSWH2006,DSOCH2007,BIRS2007}.
Here, we extend the basic idea of Hamiltonian characterization to two
coupled qubits with an unknown generic internal Hamiltonian and control
Hamiltonian response.  

This paper is organized as follows.  In Section~\ref{sec:2} we discuss
the basic principles of Hamiltonian tomography for a two-qubit system
with a fixed but unknown Hamiltonian, assuming only the ability to
measure the system at specific times in a fixed measurement basis, but
no control or \textit{a priori} knowledge of the system.  We also
deliberately exclude the ability to perform local operators on either
qubit, or the ability to initialize the system in states other than the
measurement basis states. Our approach differs in this regard from
related work on two-qubit Hamiltonian identification using concurrence
spectroscopy~\cite{Jared,Simon} or optimal experiment
design~\cite{YSKB2008}.  These approaches may be preferable for certain
types of systems but have some limitations as they presume the single
qubit dynamics can be fully characterized independently of the
inter-qubit coupling, which is required to prepare the two-qubit system
in superposition states by applying local rotations.  Using concurrence
also limits us to reconstructing the non-local part of the two-qubit
Hamiltonian.  Thus, this approach may be well-suited for some systems
e.g., with weak-coupling and non-local interaction Hamiltonians, but
may be problematic for other systems.  The approach in this paper should
be seen as complementary to these works.

In Section~\ref{sec:3} we discuss how to extract the relevant system
parameters from the noisy measurement data, accurately and robustly.
The difficulty of this task is greatly magnified compared to the single
qubit case due to the number of parameters to be determined, as well as
the increased signal complexity.  A na{\"i}ve approach using straightforward
least-squares error minimization failed completely when applied to noisy
data from simulated experiments.  The power spectrum of the signal
(which was sufficient for the single qubit case) is still useful, but no
longer an optimal frequency estimator in the presence of multiple
frequencies, and obtaining accurate estimates of the amplitudes of
different frequency components is very difficult.  For these reasons
Bayesian analysis is employed to determine the signal parameters, which
is shown to result in significant improvements in the accuracy and
robustness of the procedure.

In Section~\ref{sec:4} we show how to reconstruct the total Hamiltonian,
or more precisely, its matrix representation with respect to the fixed
measurement basis, from the estimated parameters.  Unlike the single
qubit case, calculating the $16$ matrix elements of the two-qubit
Hamiltonian from the $214$ parameters estimated from the $16$ measured
signals is non-trivial, and requires several optimization steps, from
identifying the most likely level structure from the set of transitions
frequencies, to determining the magnitudes and phases of the Hamiltonian
matrix elements that provide the best fit with the estimated
parameters. The analysis also shows that the fixed Hamiltonian can be
determined only up to a global phase and sign, as well as three phases,
which define $\U(1)$ transformations of the measurement basis states.
If there are no other measurements or control available then these
$\U(1)$ transformations of the basis states have no observable effect.
Modulo these unobservable parameters, we demonstrate that we can
reconstruct the overall Hamiltonian with very good accuracy from
noisy data.

In Section~\ref{sec:5} we consider the more general case of control
Hamiltonian tomography.  In particular, we are interested in
characterizing Hamiltonians $H=H(\vec{f})$ that depend on a number of
external parameters $\vec{f}=(f_1,\ldots,f_M)$ that can be varied
experimentally, such as voltages applied to certain gate electrodes that
allow us to vary confinement potentials, tunneling rates etc.  By
varying these parameters over time we can engineer complicated effective
Hamiltonians and efficiently achieve a wide range of control tasks from
quantum state preparation to gate implementation~\cite{some_examples}
using powerful optimal control techniques~\cite{OCT1,OCT2}.
However, effective control requires knowledge of the dependence of the
Hamiltonian on these parameters $H(\vec{f})$. When applying different
Hamiltonians, the previously unobservable phase factors now have
practical effects and are critical for full control Hamiltonian
tomography. We show how to determine these phases, relative to a
reference Hamiltonian, using a simple two-step experiment, and how to
use this information to achieve full control-Hamiltonian tomography.

Finally, in Section~\ref{sec:6} we discuss applications of the results,
as well as future improvements and generalizations to our method.

\section{Fixed Hamiltonian Tomography}
\label{sec:2}

\begin{figure*}
\includegraphics[width=\textwidth]{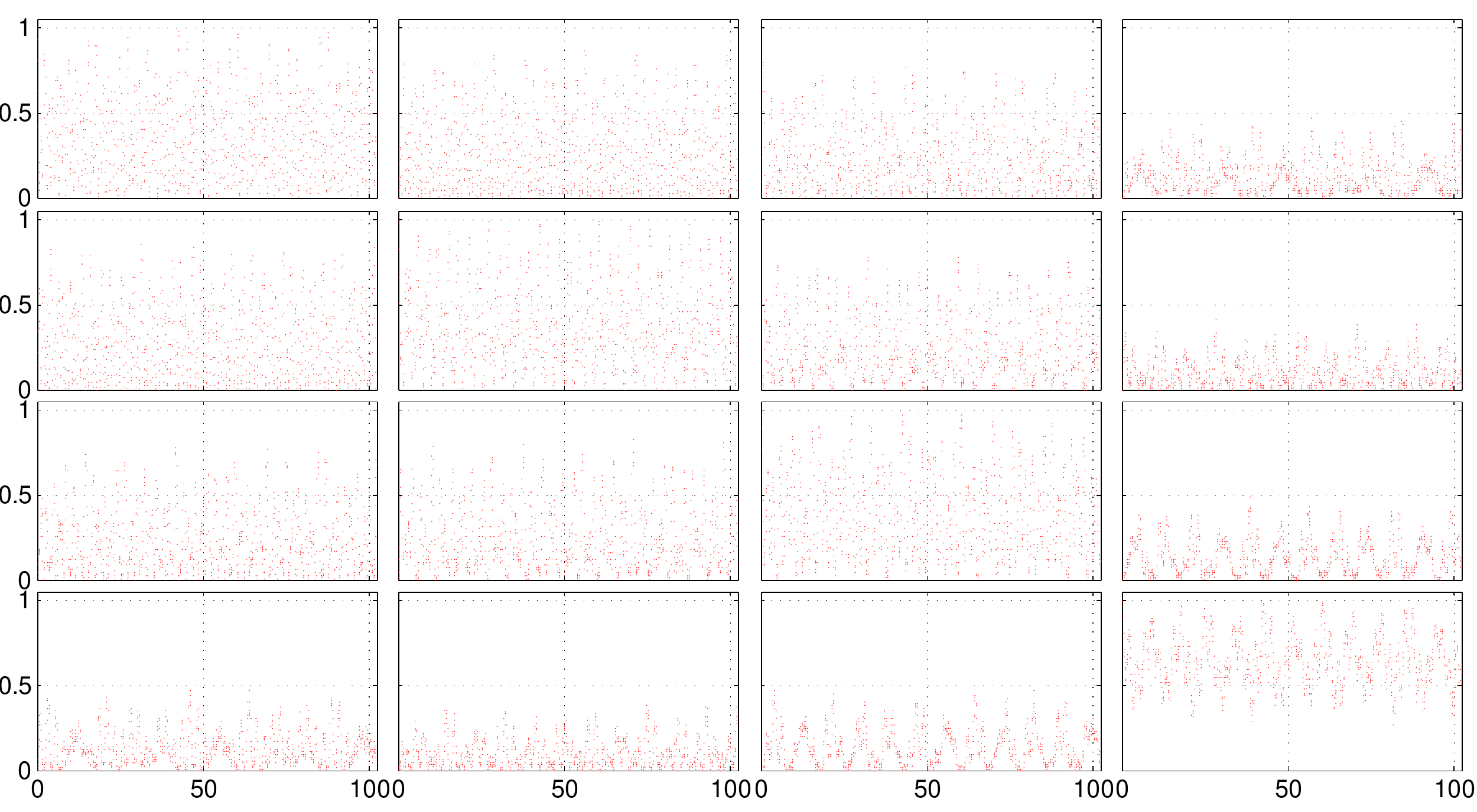}
\caption{Simulated measurements of System 1 with $2^{10}+1$ data points per
  trace sampled at $\Delta t=0.1$, signal length $T=102.4$ (arbitrary units),
  number of experiment repetitions per data point $N_e=250$. Each graph is the
  probability $p_{k\ell}(t)$ at time $t$ of detecting the system in state
  $\ket{\ell}$ ($\ell=1,2,3,4$ left to right) if initialized in state $\ket{k}$
  ($k=1,2,3,4$ top to bottom).}
\label{fig:DataSignals}
\end{figure*}

Throughout this paper we assume that we are given a two-qubit system with an
unknown Hamiltonian, and a measurement apparatus that enables us to perform a
fixed projective measurement on each qubit, including the ability to perform
effectively simultaneous measurements on both qubits~\footnote{Formally, the
  method presented will work for any four-dimensional Hamiltonian system.}.  We
denote the measurement basis states of the resulting four-outcome measurement by
$\ket{1}=\ket{00}$, $\ket{2}=\ket{01}$, $\ket{3}=\ket{10}$ and
$\ket{4}=\ket{11}$.  We then perform the following simple experiment:
\begin{enumerate} 
\item Initialize the system in one of the four measurement basis states
      $\ket{k}$ by performing simultaneous measurements on both qubits.  
\item Let the system evolve for some time $t$.
\item Perform simultaneous measurements on both qubits, projecting the 
      system back into one of the four measurement basis states. 
\end{enumerate}
By repeating this experiment many times for a fixed evolution time $t$, we can
estimate the probabilities $p_{k\ell}(t)=|\ip{\ell}{\Psi_k(t)}|^2$, where
$\ket{\Psi_k(t)}$ is the time-evolved state and $\ket{\Psi_k(0)}=\ket{k}$.  By
further repeating the experiment for different times $t_n$ for $n=0,\ldots, N-1$
we can stroboscopically capture the evolution of the probabilities
$p_{k\ell}(t)$ for $k,\ell=1,2,3,4$, yielding $16$ noisy signals as shown in
Fig.~\ref{fig:DataSignals}~\footnote{We consider only noise due to finite
  sampling for each time $t_n$ which is Poissonian. In the large $N_e$ limit,
  and for $p_{kl}\not\approx 0,1$, this tends to Gaussian noise.}.
What information about the Hamiltonian can we extract from this data, and what
is the most effective way to extract this information?

Assume the evolution of the system is governed by a fixed Hamiltonian
according to the Schr{\"o}dinger equation
\begin{equation}
  i\hbar \frac{d}{dt} \ket{\Psi(t)} = \op{H} \ket{\Psi(t)}.
\end{equation}
Expanding the Hamiltonian $\op{H}$ with respect to its orthonormal
eigenbasis $\{\ket{\xi_\nu}:\nu=1,\ldots,4\}$,
\begin{equation} 
\label{eq:expand1}
  \op{H} = \sum_{\nu=1}^4 \lambda_\nu \ket{\xi_\nu}\bra{\xi_\nu},
\end{equation}
where $\lambda_\nu$ are the (real) eigenvalues, and setting 
$\ip{k}{\xi_\nu} = r_{k\nu} e^{i\phi_{k\nu}}$ we obtain
\begin{equation}
  \bra{\ell}\op{H}\ket{k} 
  = \sum_{\nu=1}^4 \lambda_\nu \ip{\ell}{\xi_\nu}\ip{\xi_\nu}{k}
  = \sum_{\nu=1}^4 \lambda_\nu r_{\ell\nu} r_{k\nu} e^{i(\phi_{k\nu}-\phi_{\ell\nu})}.
\end{equation}
Further defining $s_{k\ell;\nu} = r_{k\nu} r_{\ell\nu}$,
$\delta_{k\ell;\nu} = \phi_{k\nu}-\phi_{\ell\nu}$ and
$\Delta_{k\ell;\mu\nu} = -\delta_{k\ell;\mu}+\delta_{k\ell;\nu}$ we obtain
\begin{equation}
  \bra{\ell}\op{H}\ket{k} 
   = \sum_{\nu=1}^4 \lambda_\nu s_{k\ell;\nu} e^{i \delta_{k\ell;\nu}}
   = e^{i\delta_{k\ell;1}} \sum_{\nu=1}^4 \lambda_\nu 
                             s_{k\ell;\nu} e^{i\Delta_{k\ell;1\nu}},
\end{equation}
where the phase terms satisfy $\delta_{\ell k;1}=-\delta_{k\ell;1}$ and
\begin{subequations}
\begin{align}
   \delta_{23;1} &= \delta_{13;1}-\delta_{12;1}, \\
   \delta_{24;1} &= \delta_{14;1}-\delta_{12;1}, \\ 
   \delta_{34;1} &= \delta_{14;1}-\delta_{13;1}.
\end{align}
\end{subequations}

If the system is initialized in one of the measurement basis states
$\ket{\Psi_k(0)}=\ket{k}$, its time-evolved state $\ket{\Psi_k(t)}$ under the
action of $\op{H}$ is given by
\begin{equation}
  \ket{\Psi_k(t)} 
   = \sum_{\nu=1}^4 e^{-i\lambda_{\nu} t} \ket{\xi_\nu} \ip{\xi_\nu}{k} 
\end{equation}
and since $\ip{\xi_\nu}{\xi_\mu} =\delta_{\nu\mu}$, its projection onto
the measurement basis state $\ket{\ell}$ at time $t$ is
\begin{equation}
  \ip{\ell}{\Psi_k(t)}
  = \sum_{\nu=1}^4 e^{-i\lambda_\nu t} \ip{\ell}{\xi_\nu} \ip{\xi_\nu}{k} \\
  = \sum_{\nu=1}^4 s_{k\ell;\nu} e^{-i(\lambda_\nu t -\delta_{k\ell;\nu})}.
\end{equation}
Hence, the probability $p_{k\ell}(t)=|\ip{\ell}{\Psi_k(t)}|^2$ of the
outcome $\ket{\ell}$ for a projective measurement of $\ket{\Psi_k(t)}$ is
\begin{multline*}
  \left[\sum_{\nu=1}^4 s_{k\ell;\nu} e^{-i(\lambda_\nu t-\delta_{k\ell;\nu})}\right] 
  \left[\sum_{\mu=1}^4 s_{k\ell;\mu} e^{ i(\lambda_\mu t-\delta_{k\ell;\mu})}\right]\\
  = \sum_{\mu=1}^4 s_{k\ell;\mu}^2 + 2\sum_{\nu>\mu}
     s_{k\ell;\mu} s_{k\ell;\nu} \cos(\omega_{\mu\nu}t - \Delta_{k\ell;\mu\nu}),
\end{multline*}
where $\omega_{\mu\nu}=\lambda_\nu-\lambda_\mu$, and using
$\cos(a-b)=\cos(a)\cos(b)+\sin(a)\sin(b)$,
\begin{equation}
\label{eq:probs}
 p_{k\ell}(t) = c_{k\ell} + 2\sum_{\nu>\mu} 
                   a_{k\ell;\mu\nu} \cos(\omega_{\mu\nu}t) 
                 + b_{k\ell;\mu\nu} \sin(\omega_{\mu\nu}t)
\end{equation}
where the coefficients are
\begin{subequations}
\label{eq:coeff}
\begin{align}
  a_{k\ell;\mu\nu} &= s_{k\ell;\nu} s_{k\ell;\mu} \cos(\Delta_{k\ell;\mu\nu}) \\
  b_{k\ell;\mu\nu} &= s_{k\ell;\nu} s_{k\ell;\nu} \sin(\Delta_{k\ell;\mu\nu}) \\
  c_{k\ell}        &= \textstyle\sum_\nu s_{k\ell;\nu}^2. 
\end{align}
\end{subequations}

Eq.~(\ref{eq:coeff}) shows that the observed dynamics are completely
determined by the transition frequencies $\omega_{\mu\nu}$, the phase
differences $\Delta_{k\ell;\mu\nu}$ and the (real) coefficients
$a_{k\ell;\mu\nu}$, $b_{k\ell;\mu\nu}$ and $c_{k\ell}$, from which we
can reconstruct the Hamiltonian $\tilde{H}$ defined by
\begin{equation}
  \label{eq:tildeH}
  \bra{\ell}\tilde{H}\ket{k} 
   = \sum_{\nu=1}^4 \tilde{\lambda}_{\nu} s_{k\ell;\nu} e^{i\Delta_{k\ell;1\nu}},
\end{equation}
where
$\tilde{\lambda}_\nu=\omega_{1\nu}-\frac{1}{4}(\omega_{12}+\omega_{13}+\omega_{14})$,
which is related to the actual Hamiltonian $H$ by
\begin{equation}
  \op{H} = D^\dag \tilde{H} D + \mbox{const.\ } \ONE,
\label{eq:gauge}
\end{equation}
where $D=\diag(1,e^{i\delta_{12}},e^{i\delta_{13}},e^{i\delta_{14}})$. The last
term is simply a global energy shift which has no observable consequences in
general. The diagonal operator $D$ represents the $\U(1)$ degree of freedom for
redefining each measurement basis state. With only a single constant
Hamiltonian, and preparation and measurement in a single fixed basis only, we
cannot completely determine the Hamiltonian.

\section{Parameter Estimation}
\label{sec:3}

The first task is to analyze the measurement traces Eq.~(\ref{eq:probs}) and extract
signal parameters Eq.~(\ref{eq:coeff}) and the frequencies $\omega_{\mu\nu}$. For
convenience we label the transition frequencies of the system $\omega_m$ for
$m=1,\ldots,6$, assuming $\omega_{m+1}>\omega_m>0$, and define the vectors
$\gvec{\omega}=(\omega_1,\ldots,\omega_6)$, $\vec{a}_{k\ell}=(a_{k\ell;m})$ and
$\vec{b}_{k\ell}=(b_{k\ell;m})$ for $k,\ell=0,1,2,3$.  The first step towards
identifying the Hamiltonian $\tilde{H}$ is to extract the six transition
frequencies $\gvec{\omega}$ and $13$ linear coefficients $\vec{a}_{k\ell}$,
$\vec{b}_{k\ell;m}$, and $c_{k\ell}$ for each of the 16 signals.  Although there
are $6+13\times 16=214$ parameters, the problem would be relatively simple if
$p_{k\ell}(t)$ was known with infinite precision for a set of sample times
$t_n$.  In practice, the accuracy of $p_{k\ell}(t_n)$ is limited by noise, in
our case projection noise due to the finite number of repetitions $N_e$, which
renders the problem one of parameter estimation for a harmonic signal with
multiple frequencies and phases. Problems of this type are common in engineering
from acoustics to image processing, and many techniques have been developed, but
our parameter estimation problem is non-trivial due to the large number of
parameters involved.

According to Eq.~(\ref{eq:coeff}) the traces $p_{k\ell}(t)$ should be
linear combinations of the $13$ basis functions 
$g_{2m-1}(t)=\cos(\omega_m t)$, $g_{2m}(t)=\sin(\omega_m t)$ for
$m=1,\ldots,6$, and $g_{13}(t)=1$, i.e.,
\begin{equation}
   p_{k\ell}(t_n) = 
   \sum_{m=1}^6 a_{k\ell,m} g_{2m-1}(t) + b_{k\ell,m} g_{2m}(t) + c_{k\ell} 
\end{equation}
and our objective is to find parameters $\omega_m$, $a_{k\ell;m}$,
$b_{k\ell;m}$ and $c_{k\ell}$ that maximize the likelihood of the
measured data.  Setting
$\vec{d}_{k\ell}=(d_{k\ell;1},\ldots,d_{k\ell;N})$, where $d_{k\ell;n}$
denotes the \emph{approximate} value of $p_{k\ell}(t_n)$ derived from
the measurement data, one way to proceed is to try to fit the parameters
to minimize the squared $L^2$-norm of the error
\begin{equation}
  \sum_{k,\ell} \norm{\vec{e}_{k\ell}}_2^2 =
  \sum_{k,\ell} \norm{\vec{p}_{k\ell}-\vec{d}_{k\ell}}_2^2,
\end{equation}
where $\vec{p}_{k\ell}=(p_{k\ell;1},\ldots,p_{k\ell,N})$ with
$p_{k\ell;n}=p_{k\ell}(t_n)$ and $\norm{\vec{e}}_2^2=\sum_{n=1}^N e_n^2$
as usual.  However, for problems with a large number of noisy data
points and a large number of parameters, as in our case, finding a
solution close to the (unknown) global minimum of the error using
brute-force optimization over all system parameters at once is difficult
at best. We tested this strategy and in most cases achieved only poor
results.

Instead of minimizing the global error, we can alternatively try to
maximize the related likelihood function
\begin{equation}
 L(\vec{a}_{k\ell},\vec{b}_{k\ell},c_{k\ell},\gvec{\omega},\sigma)  
  = \prod_{k,\ell=1}^4 \sigma_{k\ell}^{-N} \exp\left[
  -\frac{\norm{\vec{p}_{k\ell}-\vec{d}_{k\ell}}_2^2}{2\sigma_{k\ell}^2}\right].
\end{equation}
Note that we have implicitly assumed here that the signals $p_{k\ell}(t)$ are
independent and subject to Gaussian white noise with variance
$\sigma_{k\ell}^2$, assumptions that are not strictly valid in our
case. Hermitian symmetry of the Hamiltonian requires
$\vec{a}_{k\ell}=\vec{a}_{\ell k}$ and $\vec{b}_{k\ell}=-\vec{b}_{\ell k}$ but
we will enforce this symmetry later by averaging the estimated coefficients
\begin{subequations}
\label{eq:ab_symmetrization}
\begin{align}
  \vec{a}_{k\ell} &\mapsto \frac{1}{2} (\vec{a}_{k\ell}+\vec{a}_{\ell k}),\\
  \vec{b}_{k\ell} &\mapsto \frac{1}{2} (\vec{b}_{k\ell}-\vec{b}_{\ell k}).
\end{align}
\end{subequations}
The Gaussian noise model is not strictly valid either; if the measurements are
projection-noise limited then a Poissonian error model would be more accurate,
but we shall see that this is nonetheless a good approximation.

The main advantage of the latter formulation is that we can eliminate the
explicit dependence on the linear coefficients $\vec{a}_{k\ell}$,
$\vec{b}_{k\ell}$, $c_{k\ell}$ and the noise variances $\sigma_{k\ell}$ by
integration over suitable priors to obtain an explicit expression for the
probability of a particular model given the observed data $\vec{d}_{k\ell}$ that
depends only on the six transition frequencies $\gvec{\omega}$, rather than the
$>200$ parameters in the full model.  Following standard Bayesian
analysis~\cite{88Bretthorst} we obtain
\begin{equation}
  P(\gvec{\omega}|\vec{d}) \propto \prod_{k,\ell=1}^4 
  \left[
1-\frac{13\ave{\vec{h}_{k\ell}^2}}{N \ave{\vec{d}_{k\ell}^2}}
\right]^{(13-N)/2},
\end{equation}
where the averages are defined by
\begin{subequations}
\begin{align}
  \ave{\vec{d}_{k\ell}^2} &=\frac{1}{N}\sum_{n=1}^N d_{k\ell;n}^2, \\
  \ave{\vec{h}_{k\ell}^2} &=\frac{1}{13}\sum_{m=1}^{13} h_{k\ell;m}^2.
\end{align}
\end{subequations}
The components $h_{k\ell;m}$ are essentially the orthogonal projections
of the data onto a set of orthonormal basis vectors $H_m(t_n)$
\begin{equation}
  h_{k\ell;m} = \sum_{n=1}^N H_m(t_n) d_{k\ell;n}.
\end{equation}
The orthonormal basis vectors are derived from the (non-orthogonal)
basis functions $g_m(t)$ defined above, evaluated at the respective
sample times $t_n$, via
\begin{equation}
  H_m(t_n) = \frac{1}{\sqrt{\alpha_m}} \sum_{m'=1}^{13} e_{m' m} g_{m'}(t_n),
\end{equation}
where $e_{m'm}$ is a $13\times 13$ matrix whose columns $\vec{e}_m$ are
the normalized eigenvectors --- $G \vec{e}_m = \alpha_m \vec{e}_m$ ---
of the $13 \times 13$ matrix $G=(G_{m_1 m_2})$ with
\begin{equation}
  G_{m_1 m_2} = \sum_{n=1}^N g_{m_1}(t_n) g_{m_2}(t_n).
\end{equation}
The objective is to find $\gvec{\omega}$ that maximizes
$P(\gvec{\omega}|\vec{d}_{k\ell})$, or equivalently, the log-likelihood function
\begin{equation}
 \label{eq:logP}
  \log_{10} P(\gvec{\omega}|\vec{d}_{k\ell}) 
  = \frac{13-N}{2} \sum_{k,\ell=1}^4 
\log_{10} 
   \left[ 1 - \frac{13 \ave{\vec{h}_{k\ell}^2}}{N \ave{\vec{d}_{k\ell}^2}} \right].
\end{equation}

\begin{figure}
\includegraphics[width=\columnwidth]{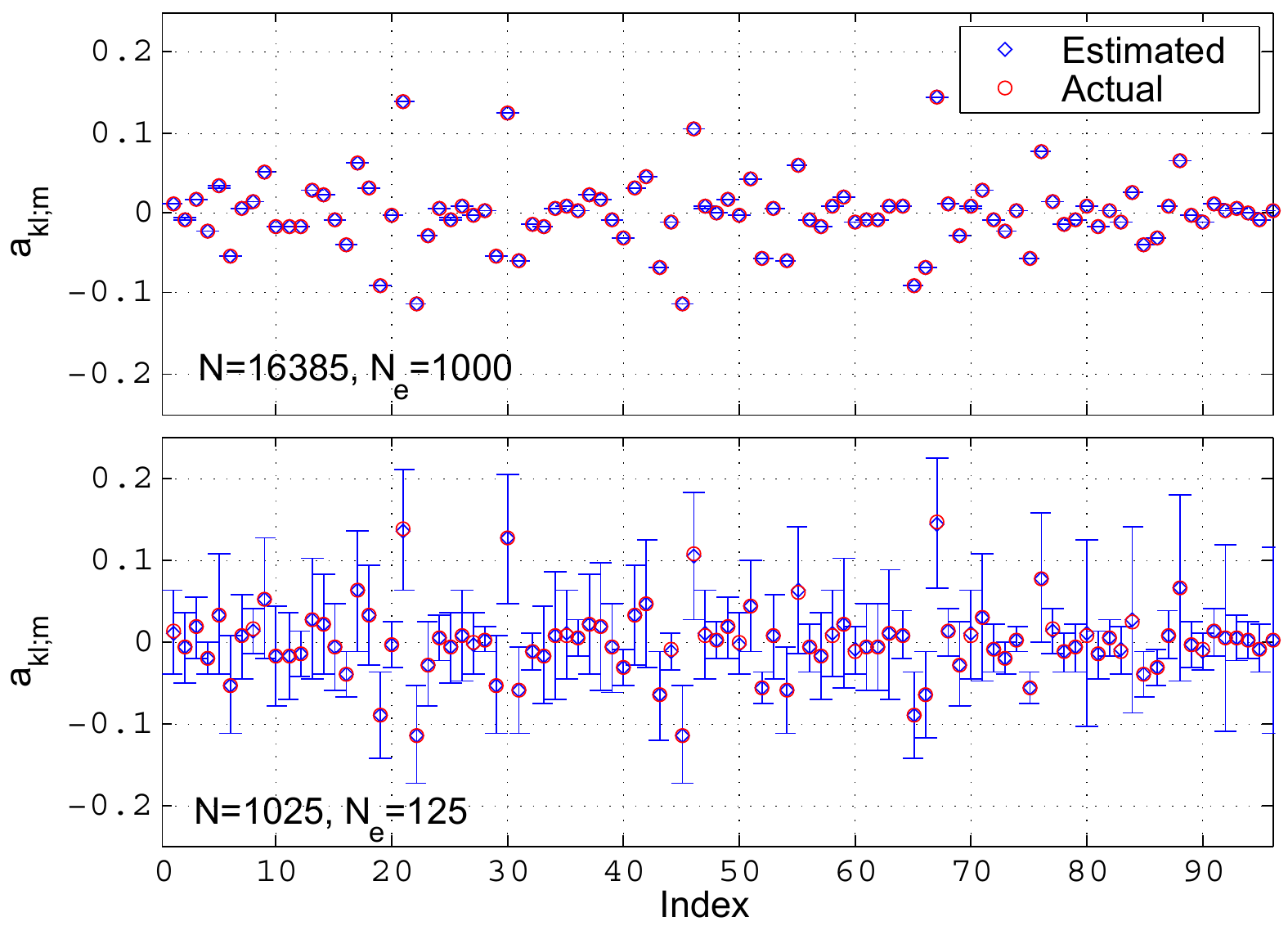}
\caption{Estimated and actual values of the coefficients $a_{k\ell;m}$ with
  estimated error-bars of System 1. When both $N$ and $N_e$ are large (top) the
  error-bars are nearly invisible, and the estimated and actual values are
  almost indistinguishable.  When $N$ and $N_e$ are both small (bottom), the
  error-bars are significantly larger, mainly due to increased noise variances
  $\sigma_{k\ell}^2$; yet the actual and estimated values for the coefficients
  are still almost indistinguishable. This suggests that the estimated
  coefficients are in fact much more accurate than the uncertainty estimates
  suggest.}
\label{fig:CoeffUncert}
\end{figure}

Note that $N$ and $\ave{\vec{d}_{k\ell}}$ are constants, while
$\vec{h}_{k\ell}$ indirectly depends on $\gvec{\omega}$ via the
basis functions $g_m(t)$.  It can be shown that the corresponding
optimal coefficients are
\begin{subequations}
\label{eq:optcoeffs}
  \begin{align}
    \vec{a}_{k\ell} &=
    \left(\ave{x_{k\ell;1}},\ave{x_{k\ell;3}},\ldots,\ave{x_{k\ell;11}}
    \right), \\
    \vec{b}_{k\ell} &=
    \left(\ave{x_{k\ell;2}},\ave{x_{k\ell;4}},\ldots,\ave{x_{k\ell;12}}
    \right), \\
    c_{k\ell} &= \ave{x_{k\ell;13}},
\end{align}
\end{subequations}
where $\ave{x_{k\ell;m}}$ is shorthand notation for the expectation values
$E(x_{k\ell;m}|\gvec{\omega},\vec{d}_{k\ell})$ of the linear coefficients of the
basis functions, given the optimal frequencies $\gvec{\omega}$ and the data
$\vec{d}_{k\ell}$.  Furthermore~\cite{88Bretthorst},
\begin{equation}
  \ave{x_{k\ell;m}} = \sum_{m'=1}^{13} \frac{e_{m m'}h_{k\ell;m'}}{\sqrt{\alpha_{m'}}}.
\end{equation}
We can similarly derive expressions for second moments
\begin{equation}
 \label{eq:2ndmoments}
  \ave{x_{k\ell;m_1}x_{k\ell;m_2}}
  -\ave{x_{k\ell;m_1}}\ave{x_{k\ell;m_2}}
  = \sigma_{k\ell}^2 \sum_{m'=1}^{13} \frac{e_{m_1 m'} e_{m_2 m'}}{\alpha_{m'}},
\end{equation}
where $\sigma_{k\ell}^2$ is the noise variance of the $(k,\ell)$th
signal, which can be approximated by its estimated expectation 
\begin{equation}
 \label{eq:sigma2}
  \ave{\sigma_{k\ell}^2} = \frac{1}{N-15}
  \left[ N \ave{\vec{d}_{k\ell}} - 13 \ave{\vec{h}_{k\ell}} \right].
\end{equation}
Note that for $m_1=m_2$ Eq.~(\ref{eq:2ndmoments}) is simply the variance
of the parameter $x_{k\ell;m}$, which gives an estimate of the
uncertainty $\Delta x_{k\ell;m}$ of the coefficient $x_{k\ell;m}$
\begin{equation}
  \Delta x_{k\ell;m}^2 \approx \mathop{\rm Var}(x_{k\ell;m}) = \ave{\sigma_{k\ell}^2}
  \sum_{m'=1}^{13} \frac{e_{m m'}^2}{\alpha_{m'}}.
\end{equation}
Fig.~\ref{fig:CoeffUncert} shows that for a sufficiently large number
of data points $N$ and experiment repetitions per data point, $N_e$,
these uncertainties can be made very small indeed.  For $N$ and/or $N_e$
small, the uncertainties are much larger, but simulations for our specific
problem suggest that the estimated values are generally still very close 
to the actual values even for small $N$ and/or $N_e$, much closer than 
the uncertainty estimates would suggest.

Although the log-likelihood function~(\ref{eq:logP}) depends explicitly only on
the six frequencies $\gvec{\omega}\in\mathbb{R}^6$ rather than the full $214$
model parameters, finding its (global) maximum is not trivial as the
log-likelihood is sharply peaked with many local extrema, and thus
computationally efficient gradient-based optimization algorithms are likely to
get trapped in local extrema if the starting point $\gvec{\omega}_0$ is chosen
randomly.  An alternative is to use global search algorithms such as pattern
search or genetic algorithms but these are computationally expensive and the
results for our problem proved inaccurate.  To circumvent this problem we adopt
a combination strategy.

\begin{figure}
\includegraphics[width=\columnwidth]{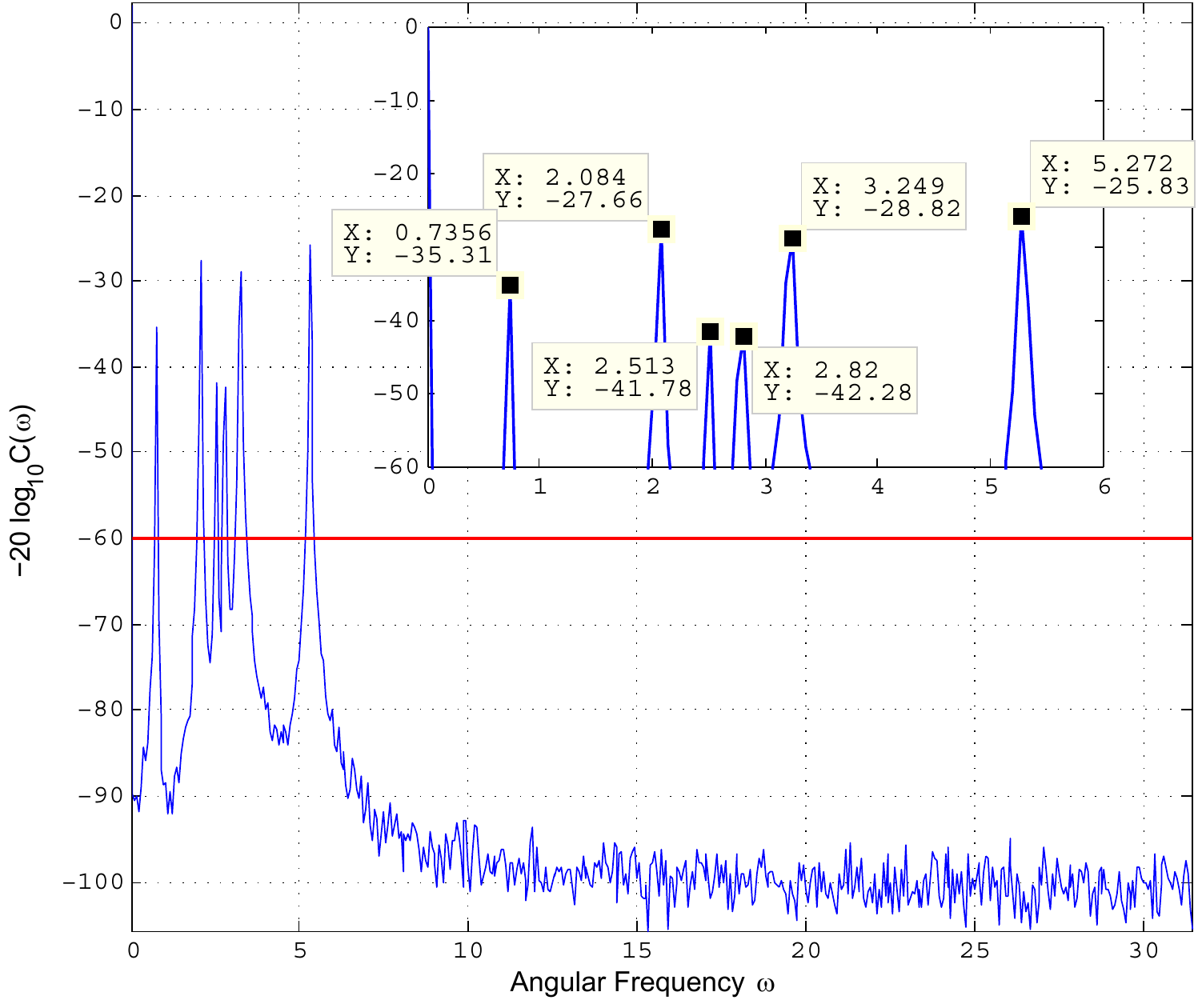} 
\caption{Power spectrum $C(\omega)$ of System 1. Although the power spectrum is
  noisy, the log-plot of $C(\omega)$ of the measured signals shown in
  Fig.~\ref{fig:DataSignals} shows six well-defined peaks for $\omega_m>0$ in
  addition to the peak at $\omega=0$.  The inset shows the filtered power
  spectrum $C(\omega)>C_0$, from which the six peaks $\omega_m$ can easily be
  identified using standard peak detection.}
\label{fig:PowerSpec}
\end{figure}

We can first estimate the resonant frequencies by looking for peaks in the power
spectra
\begin{equation} 
  C_{k\ell}(\omega) 
  = \left|\frac{1}{N}\sum_{n=1}^N d_{k\ell;n} e^{i\omega t_n} \right|^2.
\end{equation}
Using spectral filtering combined with a basic peak finding routine, we
locate (up to) six peaks $\omega_m$ in the combined power spectrum
\begin{equation}
  C(\omega)=\sum_{k,\ell=1}^4 C_{k\ell}(\omega)
\end{equation}
as illustrated in Fig.~\ref{fig:PowerSpec}, which are then used as input
$\gvec{\omega}^{(0)}=(\omega_1,\ldots,\omega_6)$ to an optimization routine
based on the BFGS quasi-Newton method with cubic line
search~\cite{BFGS1,BFGS2,BFGS3,BFGS4} to find the maximum of the
log-likelihood~(\ref{eq:logP}).  Although the discrete Fourier transform is
\emph{not} an optimal frequency estimator for a signal with multiple
frequencies, it proved generally effective in providing good starting values for
the log-likelihood optimization routine, provided that the total sampling time
(signal length) $T$ was sufficiently long to resolve the resonant peaks.

\begin{figure}
\includegraphics[width=\columnwidth]{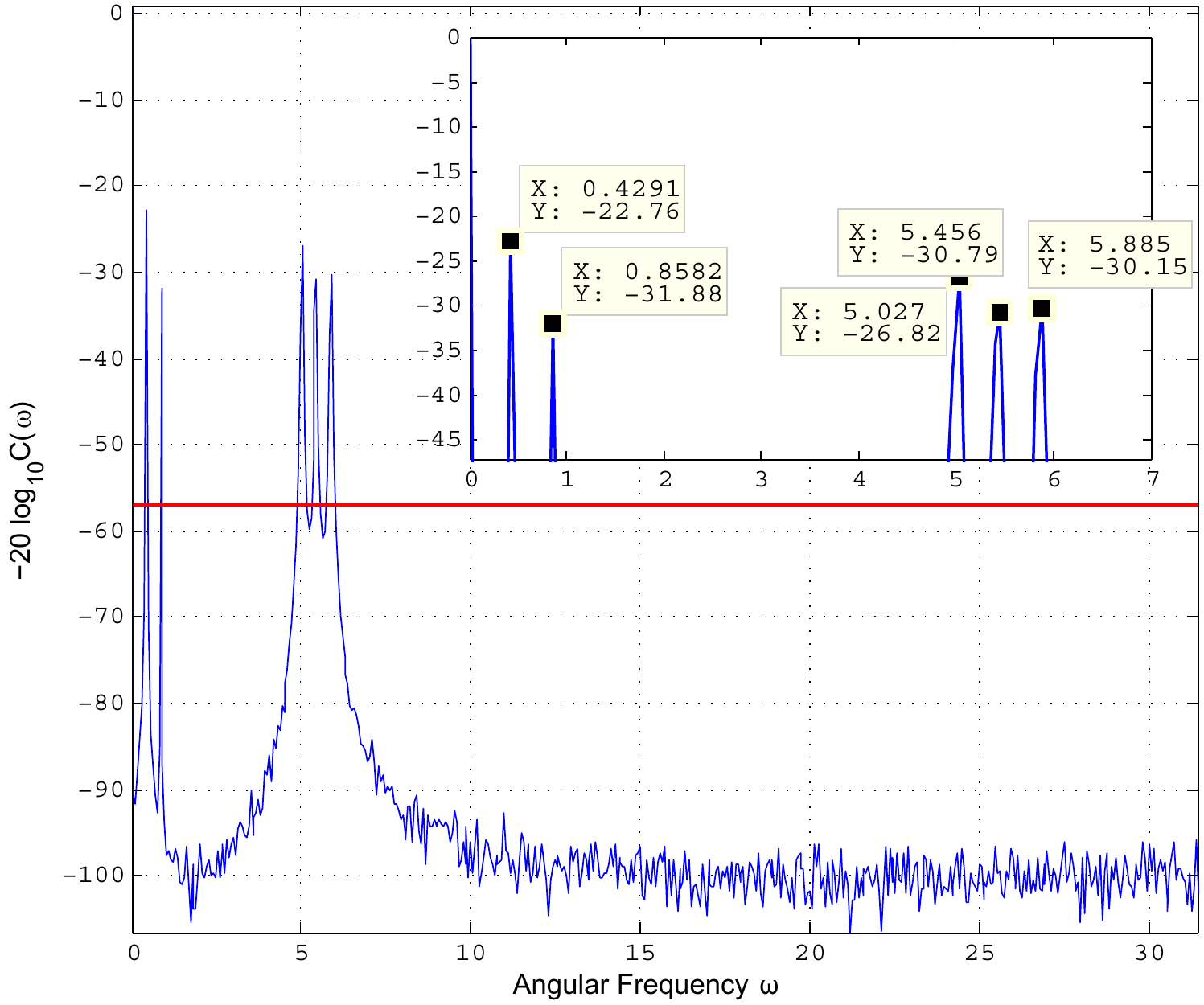} 
\caption{The power spectrum of System 12. The power Spectrum $C(\omega)$ has
  only five peaks $\omega_m>0$ in addition to the peak at $\omega=0$.  This
  could mean that the system has only five distinct transition frequencies, or
  that the measured signals are not sufficient to resolve two (closely spaced)
  transition frequencies.}
\label{fig:PowerSpec2}
\end{figure}

\begin{figure}
\includegraphics[width=\columnwidth]{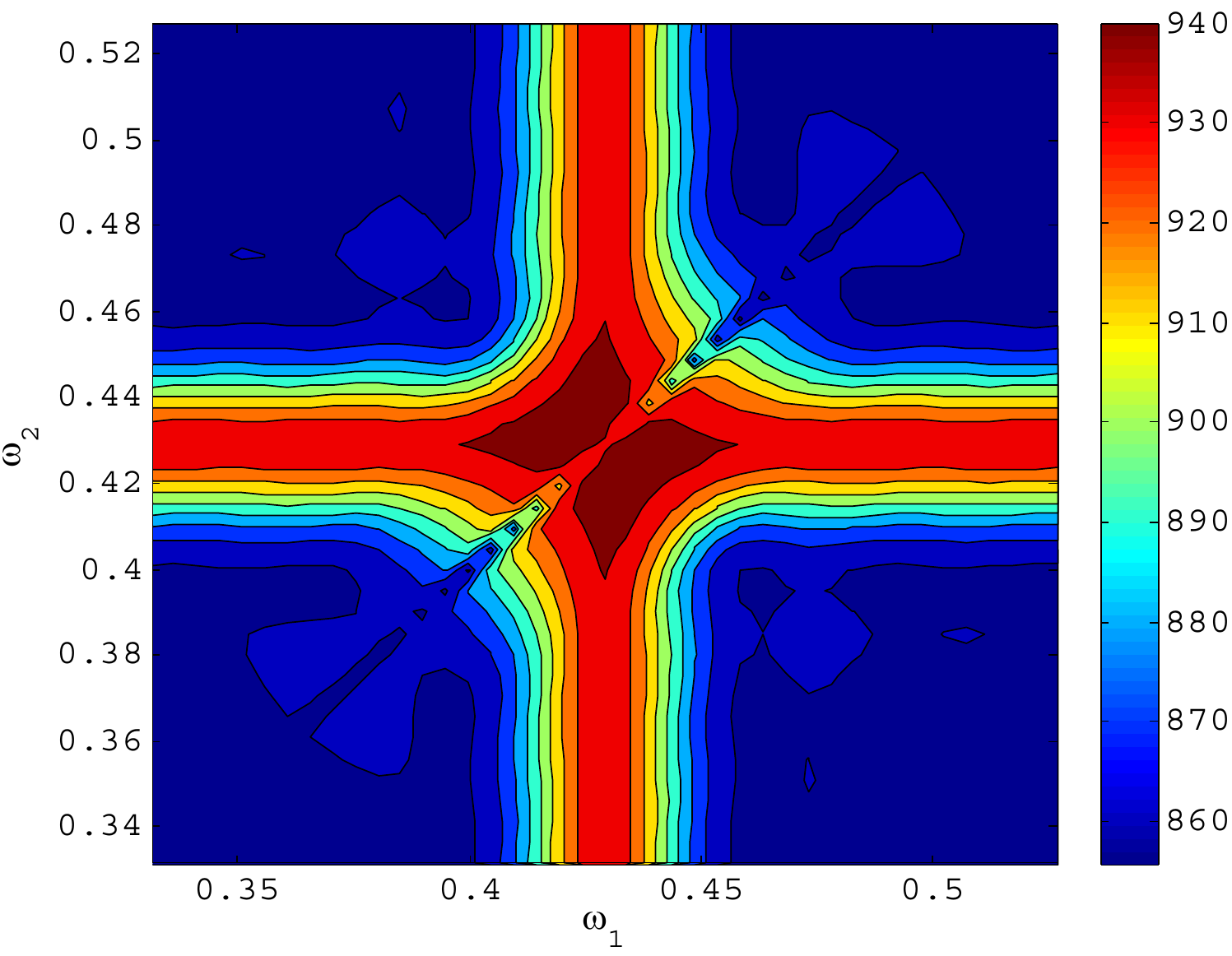}
\caption{(Color online) Log-likelihood on $I_1\times I_1$ for System 12 with
  five-peak power spectrum shown in Fig.~\ref{fig:PowerSpec2} shows symmetry
  about $y=x$ as $\log P((\omega_1,\omega_2,\ldots)|\vec{d}_{k\ell})= \log
  P((\omega_2,\omega_1,\ldots)|\vec{d}_{k\ell})$ with twin peaks for $y\neq x$
  indicating that the most probable model on this subspace of the parameter
  space is a six-frequency model.}
\label{fig:logPContour}
\end{figure}

\begin{table}
\begin{tabular}{|l|cccccc|c|}
\hline
& $\omega_1$ & $\omega_2$ & $\omega_3$ & $\omega_4$ & $\omega_5$ & $\omega_6$ & $\log P$\\\hline
$\gvec{\omega}^{(0)}$ & 0.4293& 0.8586& 4.9983& 5.4276& 5.8569&       &924.4486\\
$\gvec{\omega}^{(*)}$ & 0.4291& 0.8558& 5.0046& 5.4282& 5.8604&       &938.2960\\\hline
$\gvec{\omega}^{(1)}$ & 0.4235& 0.4323& 0.8558& 5.0046& 5.4282& 5.8604&943.3509\\\hline
$\gvec{\omega}^{(2)}$ & 0.4291& 0.7631& 0.8558& 5.0046& 5.4282& 5.8604&938.3099\\ 
$\gvec{\omega}^{(3)}$ & 0.4291& 0.8558& 5.0046& 5.1023& 5.4282& 5.8604&938.2977\\
$\gvec{\omega}^{(4)}$ & 0.4291& 0.8558& 5.0046& 5.4282& 5.5063& 5.8604&938.2993\\
$\gvec{\omega}^{(5)}$ & 0.4291& 0.8558& 5.0046& 5.4282& 5.8604& 5.9287&938.2975\\\hline
$\gvec{\omega}_{\act}$& 0.4236& 0.4322& 0.8558& 5.0046& 5.4282& 5.8604& \\\hline 
\end{tabular}
\caption{Log-likelihood for different five and six frequency models and
  actual transition frequencies for System 12 with five-peak power spectrum
  shown in Fig.~\ref{fig:PowerSpec2}.} 
\label{table1}
\end{table}

Since the frequency resolution of the power spectrum is limited by the signal
length $T$, $\Delta \omega=\frac{\pi}{T}$, if there are two or more
closely-spaced transition frequencies then it may not be possible to resolve six
peaks in the power spectrum without increasing the signal lengths significantly.
But this is generally not necessary as we can improve the frequency resolution
as follows.  Suppose there are five identifiable peaks, $\omega_1$ to
$\omega_5$, in the power spectrum, as shown in the example in
Fig.~\ref{fig:PowerSpec2}.  Then we proceed as before, using the five peak
frequencies in the power spectrum as input $\gvec{\omega}^{(0)}$ for the
optimization routine to find the most likely five-frequency model
$\gvec{\omega}^{(*)}$.  To ascertain whether there is a more probable
six-frequency model we choose an interval $I_m$ about each $\omega_m^{(*)}$,
$m=1,\ldots,5$, and investigate the log-likelihood function~(\ref{eq:logP}) on
the 2D parameter space $I_m \times I_m$, keeping the other four frequencies
fixed in each case.  E.g., for $m=1$ in the example above we find the maximum of
$\log P(\gvec{\omega}|\vec{d}_{k\ell})$ for
$\gvec{\omega}=(\omega_1,\omega_2,\omega_2^{(*)},\omega_3^{(*)},
\omega_4^{(*)},\omega_5^{(*)})$ with $(\omega_1,\omega_2) \in I_1^2$ and
$I_1=[\omega_1^{(1)}-\frac{10}{T},\omega_1^{(1)}+\frac{10}{T}]^2$ by calculating
$\log P$ on a coarse 2D grid, finding the maximum on the grid and using the
resulting $\gvec{\omega}$ as a starting point for the BFGS optimization routine
as before.  A contour plot showing the maxima in the log-likelihood on
$I_1\times I_1$ is shown in Fig.~\ref{fig:logPContour}.

We repeat this procedure for each $m$ in turn.  The results, summarized in
Table~\ref{table1}, show that the six frequency model $\gvec{\omega}^{(1)}$ is
most likely, more than the five-frequency model, and the other five
six-frequency models.  Indeed, the frequencies of the most likely six-frequency
model are very close to the actual transition frequencies of the system
simulated.  However, the relative flatness of the peak corresponding to the
global maximum of the log-likelihood function and the relatively small
differences between the likelihood of the most likely model and the less likely
models, suggests that more data would be desirable to improve the resolution of
the parameter estimates, and our confidence that the model is indeed the correct
choice. If there are fewer than five peaks in the power spectrum, the procedure
described can be iterated to sequentially resolve peaks in the power spectrum
until the most probable model has been found.

\begin{figure*}
  \includegraphics[width=\textwidth]{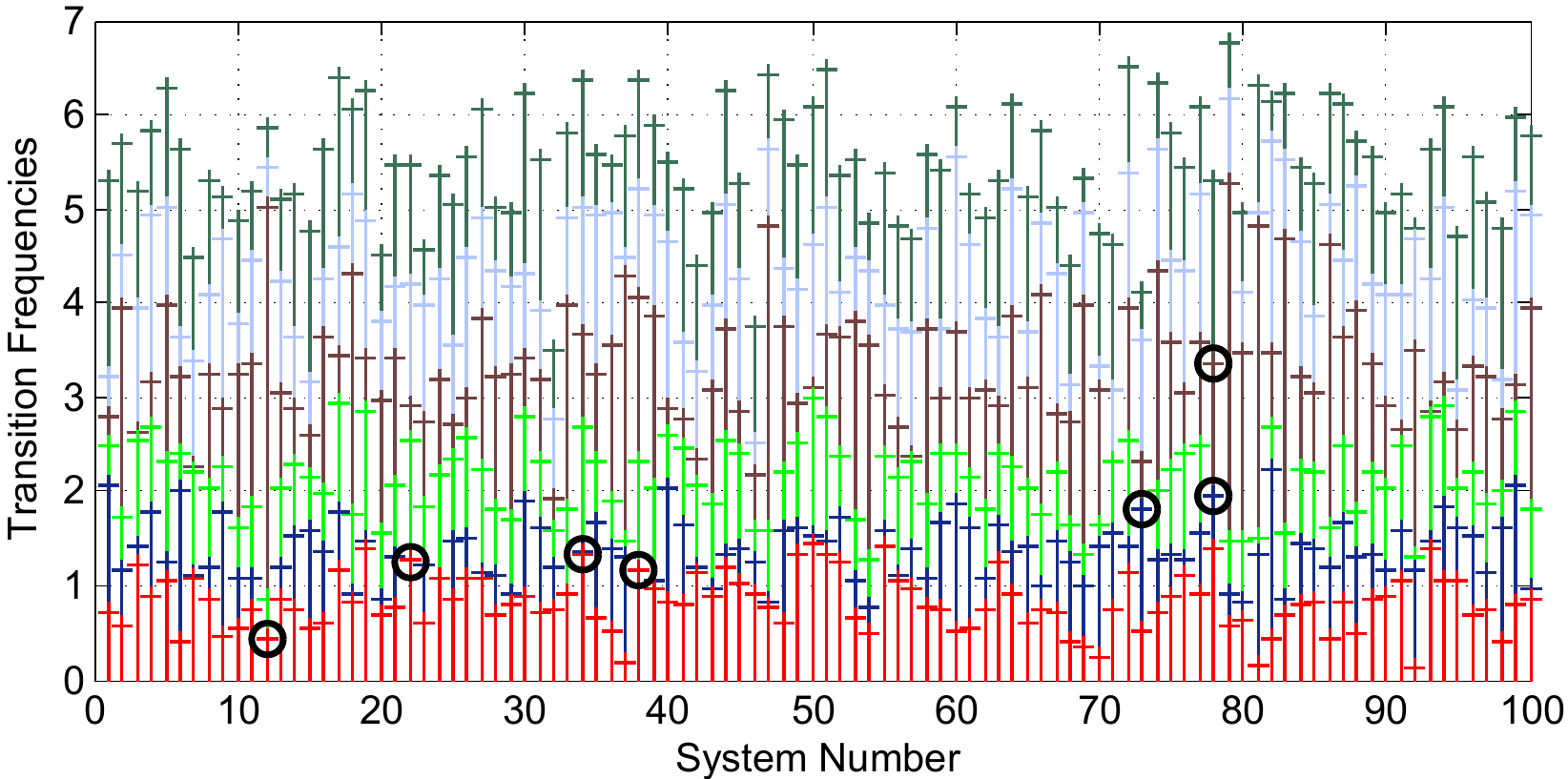}
  \caption{(Color online) The transition frequency diagram for each of the 100
    test systems shows that the transition frequencies range from 0.3 to 7, and
    there are six systems (12, 22, 34, 38, 73, 78) with two transition
    frequencies that differ by less than $0.01$ (circled), which are difficult
    to resolve, including one system (78) with two such cases.}
\label{fig:transdiag}
\end{figure*}

To test the effectiveness and accuracy of this parameter estimation
technique, we test the method for 100 randomly generated Hamiltonians,
sampled at $\Delta t=0.1$ (arbitrary units) for different signal lengths
$T=(N-1)\times\Delta t=0.1 \times 2^d$ for $d=10,11,12,13,14$ and different
levels of projection noise, with the number of measurements per data
point, $N_e\in \{125,250,500,1000\}$.  The test Hamiltonians have
transition frequencies in the range of $[0.3,7]$, and include cases with
very closely spaced transition frequencies, as shown in
Fig.~\ref{fig:transdiag}.  To assess the quality of the models found, we
calculate the transition frequencies $\omega_m$ and corresponding
parameters $\vec{a}_{k\ell}$, $\vec{b}_{k\ell}$ and $c_{k\ell}$ for each
Hamiltonian, and consider the relative errors of the parameters
identified from the noisy data with the parameter estimation technique
described.

Tables~\ref{table:err_omg_mean} and \ref{table:err_omg_med} show the
means and medians, respectively, over 100 systems, of the maximum
relative error (in percent)
\begin{equation}
  \eps_{\max}(\gvec{\omega}^{(0)}) 
  = 100 \times  \max_{m\in{1,\ldots,6}} 
    \left|1-\frac{\omega_m^{(0)}}{\omega_m}\right|
\end{equation}
of the estimated transition frequencies for each system, where $\omega_m$ 
are the exact transition frequencies.  Comparison of the errors for the 
initial frequency estimates obtained from the power spectrum, labeled
$\gvec{\omega}^{(0)}$, and the optimal values $\gvec{\omega}^{\opt}$ 
obtained by maximizing the log-likelihood shows the optimized frequencies 
are generally about two orders of magnitude more accurate than the 
estimates obtained from the power spectrum.  

\begin{table}
\begin{tabular}{|r|cccc|cccc|}
\hline
$N \backslash N_e$ & 125 & 250 & 500 & 1000 & 125 & 250 & 500 & 1000 \\\hline
16,385 & 0.093 & 0.094 & 0.094 & 0.094 & 0.0002 & 0.0002 & 0.0001 & 0.0001\\
8,193  & 0.231 & 0.226 & 0.231 & 0.231 & 0.0006 & 0.0006 & 0.0004 & 0.0003\\
4,097  & 0.432 & 0.432 & 0.432 & 0.432 & 0.0018 & 0.0019 & 0.0009 & 0.0009\\
2,049  & 0.696 & 0.685 & 0.680 & 0.685 & 0.0065 & 0.0040 & 0.0030 & 0.0024\\
1,025  & 1.646 & 1.650 & 1.646 & 1.650 & 0.0272 & 0.0184 & 0.0085 & 0.0108\\\hline
\end{tabular}
\caption{The percentage relative errors
$\ave{\eps_{\max}(\gvec{\omega}^{(0)})}$ (left) and
$\ave{\eps_{\max}(\gvec{\omega}^{\opt})}$ (right) show that the
log-likelihood optimization improves the accuracy of the frequency
estimates by at least two orders of magnitude compared to the estimates
obtained from the power spectrum.}  \label{table:err_omg_mean}
\end{table}

\begin{table}
\begin{tabular}{|r|cccc|cccc|}
\hline
$N \backslash N_e$ & 125 & 250 & 500 & 1000 & 125 & 250 & 500 & 1000 \\\hline
16,385 & 0.068 & 0.068 & 0.068 & 0.068 & 0.0001 & 0.0001 & 0.0001 & 0.0001\\
8,193  & 0.167 & 0.164 & 0.167 & 0.167 & 0.0005 & 0.0003 & 0.0002 & 0.0002\\
4,097  & 0.327 & 0.327 & 0.327 & 0.327 & 0.0011 & 0.0012 & 0.0006 & 0.0005\\
2,049  & 0.5100 & 0.493 & 0.493 & 0.493 & 0.0035 & 0.0023 & 0.0019 & 0.0011\\
1,025  & 1.164 & 1.142 & 1.164 & 1.142 & 0.0126 & 0.0089 & 0.0052 & 0.0036\\\hline
\end{tabular}
\caption{The medians of percentage relative errors
  $\eps_{\max}(\gvec{\omega}^{0})$ (left) and
  $\eps_{\max}(\gvec{\omega}^{\opt})$ (right) show the same accuracy
  improvements of the log-likelihood estimates.  Median errors lower than
  the averages indicate that the error distribution is peaked towards the origin.}
\label{table:err_omg_med}
\end{table}

\begin{table}
\begin{tabular}{|l|r|cccc|}
\hline
& $N \backslash N_e$ & 125 & 250 & 500 & 1000 \\\hline
$\ave{\eps_{\rm med}(a_{k\ell;m})}$ 
 &16,385 & 0.3825 & 0.2671 & 0.1912 & 0.1454\\
 & 8,193 & 0.5538 & 0.3598 & 0.2857 & 0.1923\\
 & 4,097 & 0.7711 & 0.5516 & 0.4075 & 0.2786\\
 & 2,049 & 1.0630 & 0.7940 & 0.5755 & 0.3762\\
 & 1,025 & 1.5817 & 1.1210 & 0.7880 & 0.5573\\\hline
$\ave{\eps_{\rm med}(b_{k\ell;m})}$ 
 &16,385 & 0.2417 & 0.1739 & 0.1174 & 0.0846 \\
 & 8,193 & 0.3333 & 0.2519 & 0.1755 & 0.1144\\
 & 4,097 & 0.4860 & 0.3470 & 0.2394 & 0.1733\\
 & 2,049 & 0.6715 & 0.5098 & 0.3436 & 0.2485\\
 & 1,025 & 1.0194 & 0.7197 & 0.4691 & 0.3523\\\hline
$\ave{\eps_{\rm med}(c_{k\ell})}$ 
 &16,385 & 0.0734 & 0.0525 & 0.0378 & 0.0279\\
 & 8,193 & 0.1002 & 0.0751 & 0.0538 & 0.0372\\
 & 4,097 & 0.1463 & 0.1037 & 0.0770 & 0.0518\\
 & 2,049 & 0.2007 & 0.1483 & 0.1148 & 0.0751\\
 & 1,025 & 0.2817 & 0.2258 & 0.1555 & 0.1047\\\hline
$\ave{\sigma^2}$ 
&16,385 & 0.0012 & 0.0006 & 0.0003 & 0.0001 \\
& 8,193 & 0.0971 & 0.0959 & 0.0953 & 0.0950 \\
& 4,097 & 0.2896 & 0.2873 & 0.2861 & 0.2855 \\
& 2,049 & 0.6763 & 0.6717 & 0.6692 & 0.6681 \\
& 1,025 & 1.4580 & 1.4487 & 1.4437 & 1.4414 \\\hline
\end{tabular}
\caption{Relative errors $\eps_{\rm med}(a_{k\ell;m})$, $\eps_{\rm
med}(b_{k\ell;m})$, and $\eps_{\rm med}(c_{k\ell})$ (in \%) and
estimated error variances $\ave{\sigma^2}$, averaged over 100
test systems for different signal length $T=0.1(N-1)$ and number of
experiment repetitions $N_e$ per data point.}  \label{table:err_a_b_c}
\end{table}

\begin{figure}
  \includegraphics[width=\columnwidth]{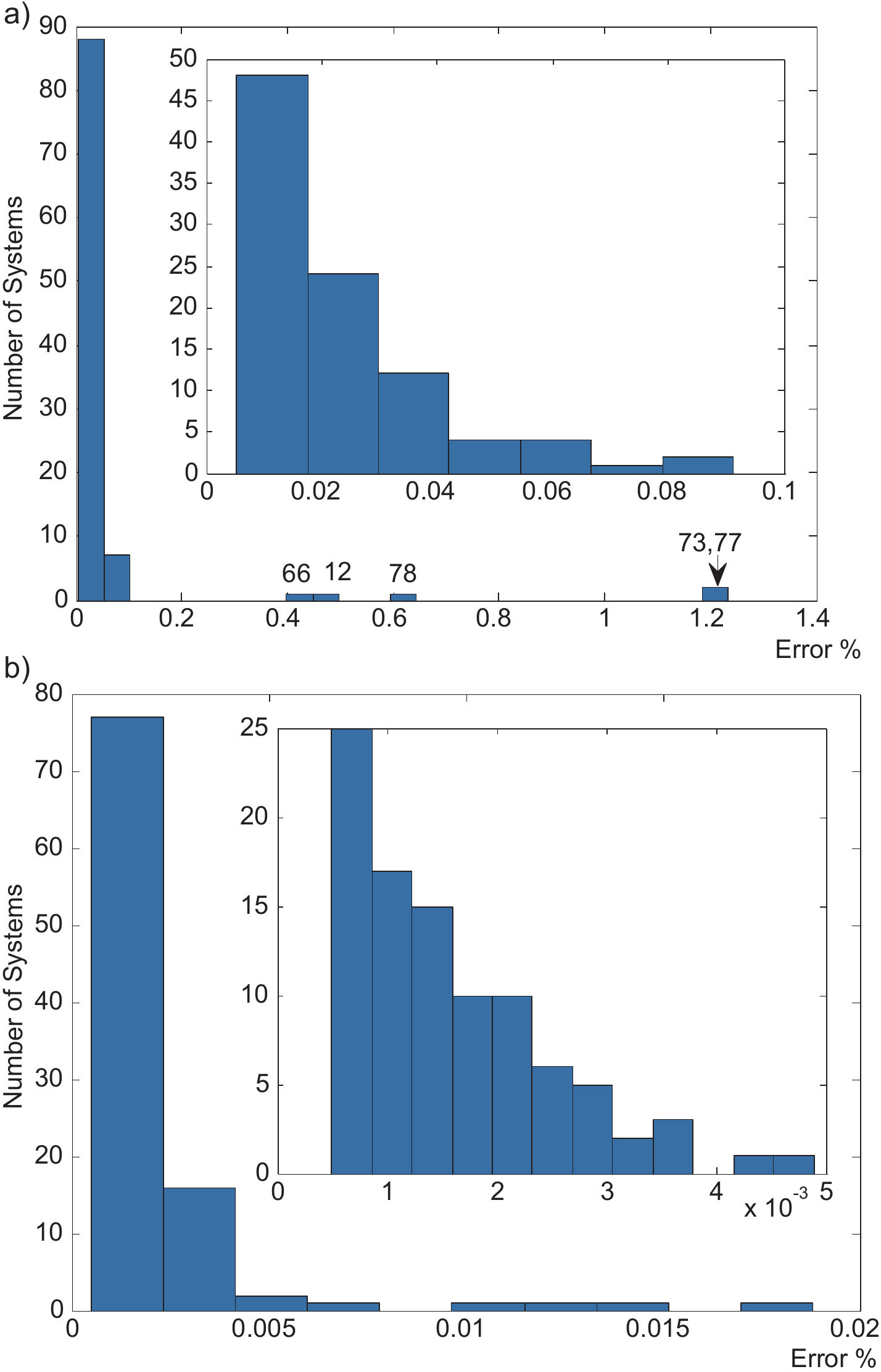}
  \caption{Histogram of the relative \% error for the test 100 systems for
    sampling numbers of a) $N=1025,N_e=125$ b) $N=16385,N_e=1000$. Inset graphs
    magnify the region around the origin showing the general distribution of
    errors which is roughly exponential. The numbers 66, 12, 78, 73 and 77 in a)
    refer to outliers systems.}
\label{fig:histo}
\end{figure}

The linear coefficients $a_{k\ell;m}$, $b_{k\ell;m}$ and $c_{k\ell}$ are then
estimated from the maximization of Eq.~(\ref{eq:logP}) and
from~Eq.(\ref{eq:optcoeffs}). Taking the median of the relative errors
\begin{equation}
  \eps_{\rm med}(a_{k\ell;m}) 
  = 100\% \times \mathop{\rm median}\limits_{k,\ell,m} 
    \left|1-\frac{a_{k\ell;m}^{\est}}{a_{k\ell;m}}\right|,
\end{equation}
where $k,\ell$ range from $1$ to $4$ and $m=1,\ldots,6$, as a general measure of
the quality of the fit, Table~\ref{table:err_a_b_c} shows that the average
errors in the coefficients $a_{k\ell;m}$, $b_{k\ell;m}$ and to a lesser extent
$c_{k\ell}$, are generally at least one order of magnitude larger than the error
in the frequency estimates. Overall the quality is still good, however, with the
(average) errors ranging from a fraction of a percent to less than $2.5$\% for
$a_{k\ell}$, and much less for $c_{k\ell}$, depending on the number of data
points $N$ and the accuracy of the data points determined by the number of
experiment repetitions per data point, $N_e$. Fig.~\ref{fig:histo} shows the
distribution of the errors for both the least and greatest number of
experiments. Apart from a few outliers, the distribution follows a roughly
exponential form with most estimates being within a fraction of a percent of the
true values, even for the least number of experimental samples.

\begin{figure}
  \includegraphics[width=\columnwidth]{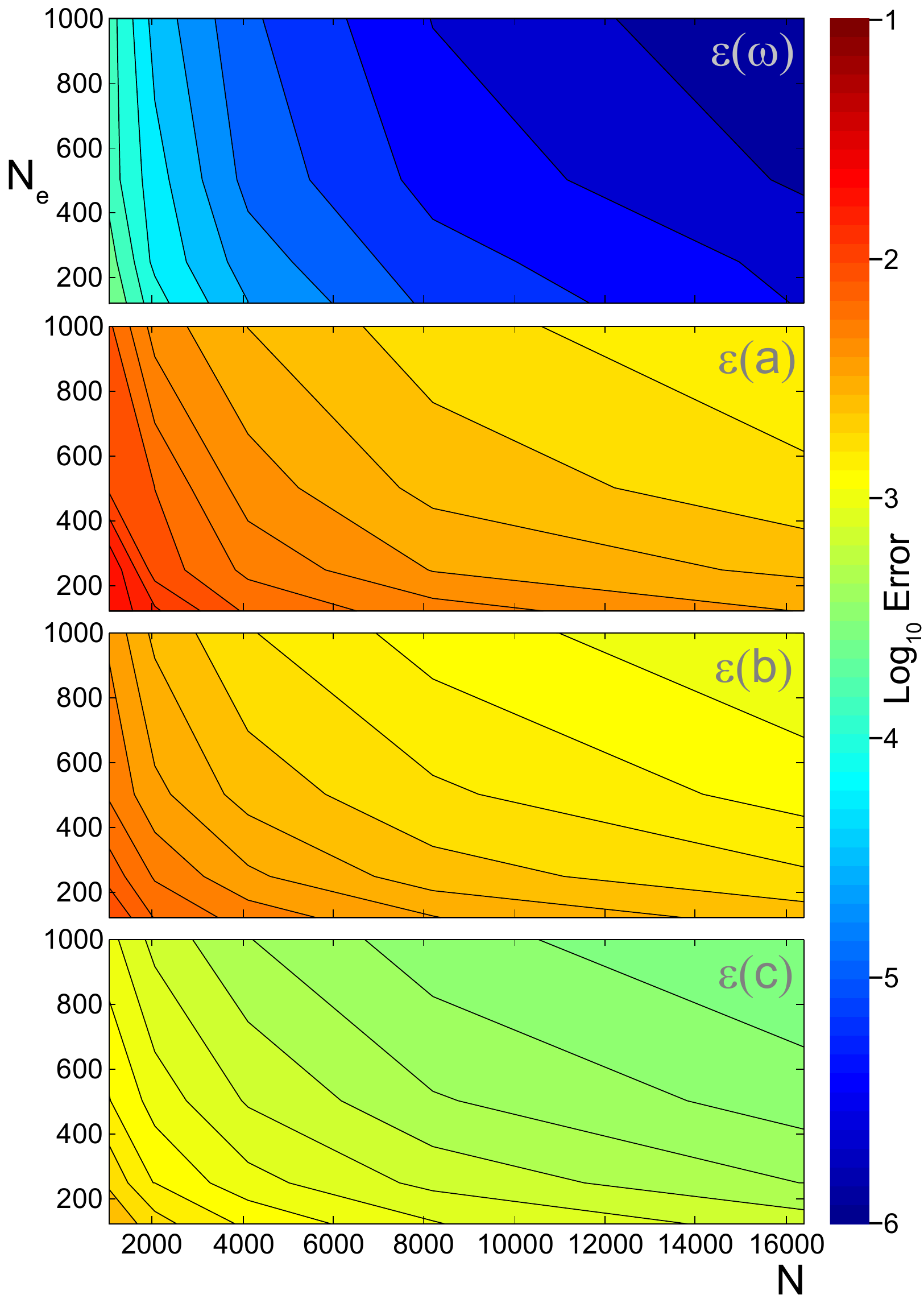}
  \caption{(Color online) Contour plots of the $\log_{10}$ mean (relative)
    errors for the frequencies $\omega$ and coefficients $a$, $b$, $c$. The
    frequencies show the smallest errors (down to $10^{-6}$), whilst the $a$
    coefficients show errors up to a few percent ($\tilde 10^{-2}$) for the
    shortest signal lengths and greatest projection noise.}
\label{fig:error_contour_plots}
\end{figure}

Table~\ref{table:err_omg_mean} shows that increasing $N_e$ and thus the accuracy
of the data points does not improve the accuracy of the initial frequency
estimates obtained from the power spectrum at all, while doubling $N$ tends to
reduce the error by more than half.  This is what we expect as once $N_e$ is
large enough to permit discrimination of the resonant peaks from the noise
floor, little is gained by increasing $N_e$.  Doubling $N_e$ does reduce the
error for the optimized frequencies obtained from our Bayesian analysis, although
if the accuracy of frequency estimates alone is considered, doubling the number
of data points in preferable to doubling $N_e$.  Increasing the accuracy (by
doubling $N_e$) is more effective in reducing the errors in the coefficients
$a$, $b$, $c$, but the contour plots in Fig.~\ref{fig:error_contour_plots} show
that the errors decrease faster with $N$, i.e., increasing the number of data
points is generally still preferable.

\section{Hamiltonian Reconstruction}
\label{sec:4}

\begin{figure*}
\includegraphics[width=\textwidth]{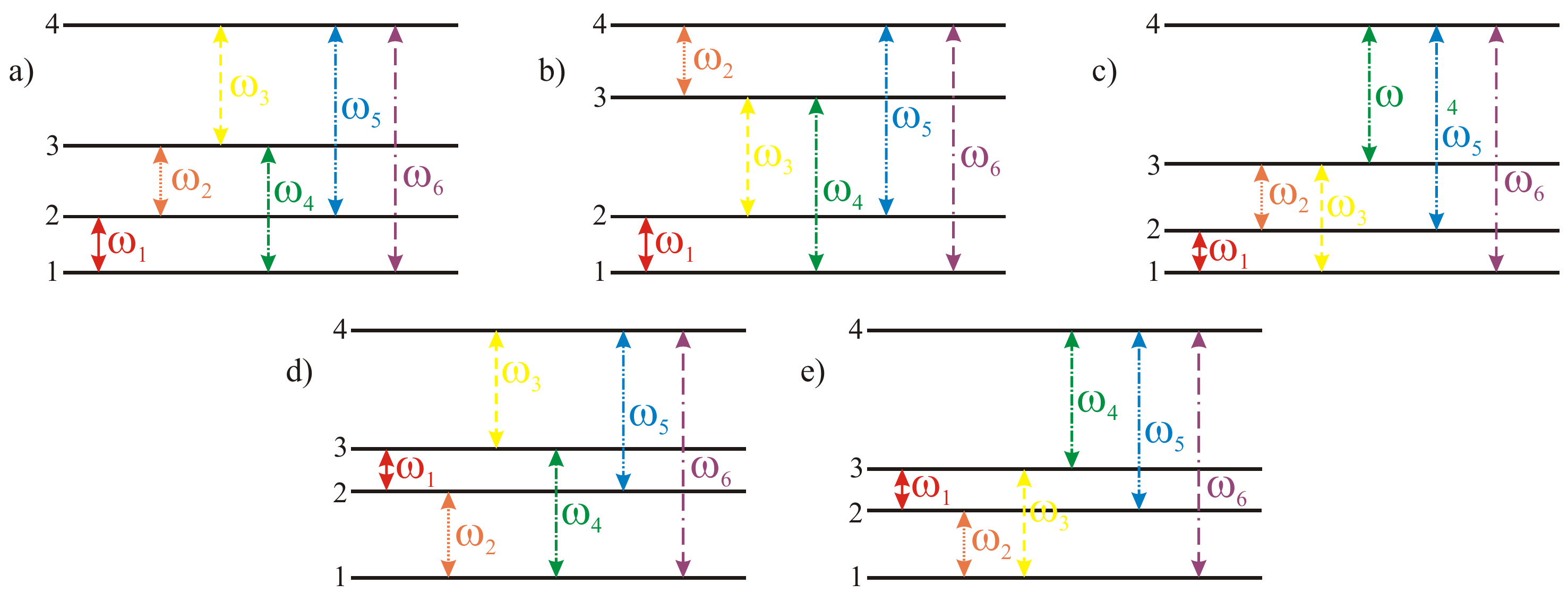}
\caption{(Color online) Possible arrangements for a generic four-level system
  with six distinct transition frequencies. Not shown are the other five
  configurations which correspond to a reflection of the energies, which merely
  flips the above level structures.}
\label{fig:levels}
\end{figure*}

Once the frequencies $\gvec{\omega}$ and amplitudes $\vec{a}_{k\ell}$,
$\vec{b}_{k\ell}$ and $c_{k\ell}$ have been extracted from the measured data
using parameter estimation, reconstructing the Hamiltonian (up to equivalence)
requires at least two further steps: identification of the resonant frequencies
with transitions $(\mu,\nu)$ between eigenstates $\ket{\xi_\mu}$ and
$\ket{\xi_\nu}$ of the system, and computation of the parameters $s_{k\ell;\nu}$
and $\Delta_{k\ell;\mu\nu}$ in Eq.~(\ref{eq:tildeH}) from the coefficients
$\vec{a}_{k\ell}$, $\vec{b}_{k\ell}$ and $c_{k\ell}$.  For a four-level system
we have three primary transitions $\{\omega_{12},\omega_{23},\omega_{34}\}$
between adjacent energy levels and three other transitions
$\{\omega_{13},\omega_{24},\omega_{14}\}$, which must satisfy
\begin{subequations}
\label{eq:freqrel}
\begin{align}
 \omega_{13} &=\omega_{12}+\omega_{23},\\
 \omega_{24} &=\omega_{23}+\omega_{34},\\
 \omega_{14} &=\omega_{12}+\omega_{23}+\omega_{34}.
\end{align}
\end{subequations}

We identify the possible level structure (up to inversion) by examining
the relationships between the frequencies.  In the generic case, i.e.,
when there are six distinct transition frequencies,
$0<\omega_1<\omega_2<\ldots<\omega_6$, it follows immediately from
Eqs~(\ref{eq:freqrel}) that $\omega_6=\omega_{14}$, and the primary
transitions are $\{\omega_1,\omega_2,\omega_6-\omega_1-\omega_2\}$.
Closer inspection shows that there are $10$ possible arrangements of the
six transition frequencies as shown in Fig.~\ref{fig:levels}, and the
exact transition frequencies $\gvec{\omega}$ must satisfy
$A_s\gvec{\omega}=\vec{0}$ for one of the following matrices
\begin{subequations}
\begin{gather}
A_1=\begin{bmatrix} 1&1&1&0&0&-1 \\ 1&1&0&-1&0&0 \\ 0&1&1&0&-1&0 \end{bmatrix}
A_2=\begin{bmatrix} 1&1&1&0&0&-1 \\ 1&0&1&-1&0&0 \\ 0&1&1&0&-1&0 \end{bmatrix}\nonumber\\
A_3=\begin{bmatrix} 1&1&1&0&0&-1 \\ 1&1&0&-1&0&0 \\ 1&0&1&0&-1&0 \end{bmatrix}
A_4=\begin{bmatrix} 1&1&0&1&0&-1 \\ 1&1&-1&0&0&0 \\ 1&0&0&1&-1&0 \end{bmatrix}\nonumber\\
A_5=\begin{bmatrix} 1&1&0&1&0&-1 \\ 1&1&-1&0&0&0 \\ 0&1&0&1&-1&0 \end{bmatrix}.
\end{gather}
\end{subequations}
Given the estimated frequencies $\gvec{\omega}^{opt}$ the most likely
case is that for which $\norm{A_s \gvec{\omega}^{opt}}_2^2$ assumes its
minimum, which should be close to $0$, and significantly smaller than
the errors for the other cases.  A larger minimum error indicates and
none of the possibilities is likely, suggesting that the system may not
be a Hamiltonian four-level system.  Similarly, if we have two cases for
which the error the close to the minimum, this would be an indication
that further data is required to resolve the ambiguity.  

\begin{figure}
\includegraphics[width=\columnwidth]{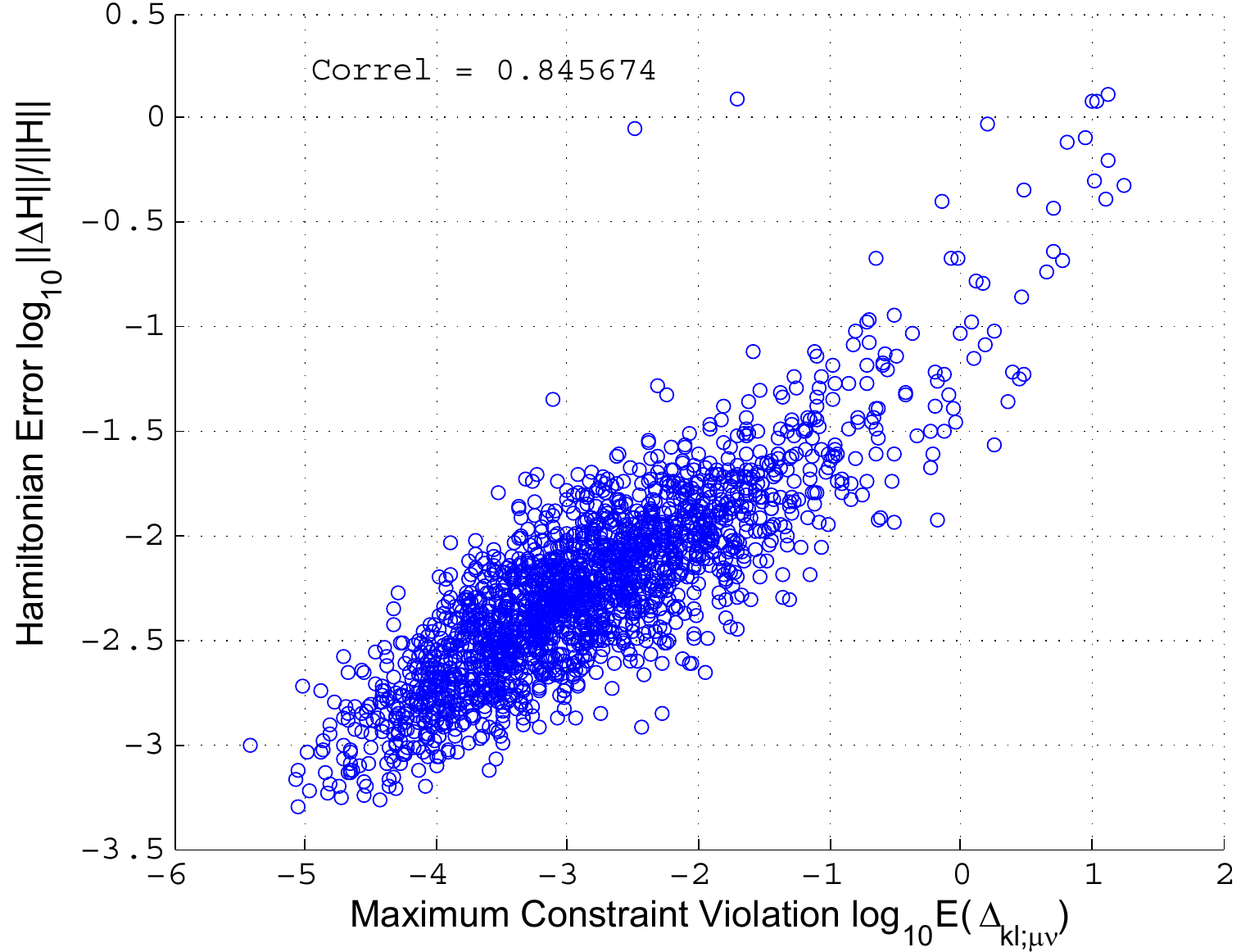}
\caption{A scatterplot of the relative errors $\norm{\Delta H}/\norm{H}$ of the
  estimated Hamiltonian with $\norm{\Delta H}$ as defined in~(\ref{eq:DeltaH})
  vs the maximum constraint violation~(\ref{eq:MaxConstraintVio}) on a log-log
  scale for our 100 systems and 20 data sets per system (total of 2000 data
  points) shows a strong correlation, suggesting that the maximum constraint
  violation (prior to refinement) is a good predictor of the accuracy of the
  estimated Hamiltonian.}  \label{fig:ErrorCorrel}
\end{figure}

Once the observed frequencies $\omega_m$ have been matched with actual
transitions $(\mu,\nu)$, we can associate the corresponding coefficients
$a_{k\ell,m}$, $b_{k\ell,m}$ for $m=1,\ldots,6$ with their respective
transitions, i.e., we have $a_{k\ell;\nu\ell}$ and $b_{k\ell;\nu\ell}$,
and determine the phase differences
\begin{equation}
 \label{eq:Delta}
  \Delta_{k\ell;\mu\nu}= \arctan(b_{k\ell;\mu\nu}, a_{k\ell;\mu\nu}),
\end{equation}
where $\arctan(b,a)$ is the four-quadrant arc tangent of $b/a$.  If the
estimated parameters are good, then the resulting $\Delta_{k\ell;\mu\nu}$
should satisfy $\Delta_{kk;\mu\nu}\approx 0$ (mod $2\pi$),
$\Delta_{k\ell;\mu\nu}\approx -\Delta_{k\ell;\mu\nu}$ (mod $2\pi$), and
\begin{subequations}
\label{eq:Delta_constraints}
\begin{align}
   \Delta_{k\ell;12}+\Delta_{k\ell;13} -\Delta_{k\ell;23} &=0 \mod\, 2\pi\\
   \Delta_{k\ell;13}+\Delta_{k\ell;14} -\Delta_{k\ell;34} &=0 \mod\, 2\pi\\
   \Delta_{k\ell;12}+\Delta_{k\ell;14} -\Delta_{k\ell;24} &=0 \mod\, 2\pi.
\end{align}
\end{subequations}
Due to the enforced symmetrization~(\ref{eq:ab_symmetrization}) of the
coefficients $\vec{a}_{k\ell}$ and $\vec{b}_{k\ell}$, the phase terms
should satisfy $\Delta_{k\ell;\mu\nu}=-\Delta_{\ell k;\mu\nu}$.  Minor 
violations of~(\ref{eq:Delta_constraints}) are to be expected, and can 
be mitigated, and the accuracy of the final reconstructed Hamiltonian 
improved by minimizing the constraint violations
$\norm{\vec{e}_{k\ell}}_2^2 =\sum_{s=1}^3 e_{k\ell;s}^2$, where
\begin{equation}
  e_{k\ell;s} = \min \{ |x_{k\ell;s}|,|x_{k\ell;s}-2\pi|,|x_{k\ell;s}+2\pi| \},
\end{equation}
with $\vec{x}_{k\ell}= A\gvec{\Delta}_{k\ell}$ and
\begin{equation}
  A = \begin{bmatrix} 
       1 & 1 & 0 & -1 & 0 & 0 \\
       1 & 0 & 0 & 0 & 1 & -1 \\
       0 & 0 & 1 & 1 & 0 & -1 
      \end{bmatrix},
\end{equation}
for $k,\ell=1,\ldots,4$ in a further refinement step, starting with the values
for $\Delta_{k\ell;\mu\nu}$ obtained from (\ref{eq:Delta}). This refinement
tries to minimize the discrepancy between the estimated signal parameters and
those expected from an underlying Hamiltonian model. It must be stressed,
however, that larger violations of the constraints are indicative of significant
errors, which may even be exacerbated by such a refinement. In fact,
Fig.~\ref{fig:ErrorCorrel} shows that there is a strong correlation between the
maximum constraint violation prior to refinement
\begin{equation}
  \label{eq:MaxConstraintVio}
  \mathcal{E}(\Delta_{k\ell;\mu\nu}) = \max_{k,\ell} \norm{\vec{e}_{k\ell}}_2^2
\end{equation}
and the relative error of the final estimated Hamiltonian.

Once the optimal values for $\Delta_{k\ell;\mu\nu}$ have been found, we
calculate the products
\begin{equation} 
\label{eq:Mkell}
  s_{k\ell;\mu}s_{k\ell;\nu} =
  a_{k\ell;\mu\nu}\cos(\Delta_{k\ell;\mu\nu})+b_{k\ell;\mu\nu}\sin(\Delta_{k\ell;\mu\nu}).
\end{equation}
Labelling the RHS of the previous equation $M_{k\ell;\mu\nu}$ and
defining the column vector $\vec{s}_{k\ell}$ and the $4\times 4$ matrix
$M_{k\ell}$
\begin{equation*}
 \vec{s}_{k\ell}= 
 \begin{pmatrix} s_{k\ell;1}\\s_{k\ell;2}\\s_{k\ell;3}\\s_{k\ell;4}
 \end{pmatrix}, \quad
  M_{k\ell} = 
 \begin{pmatrix}
  M_{k\ell;11} & \ldots & M_{k\ell;14} \\
  \vdots & \ddots & \vdots \\
  M_{k\ell;41} & \ldots & M_{k\ell;44}
 \end{pmatrix}
\end{equation*}
we can express Eqs~(\ref{eq:Mkell}) and (\ref{eq:coeff}c) as follows
\begin{equation}
 \label{eq:Ms}
   \vec{s}_{k\ell} \vec{s}_{k\ell}^T = M_{k\ell}, \qquad
   \vec{s}_{k\ell}^T \vec{s}_{\ell} = c_{k\ell}
\end{equation}
for $k,\ell=1,\ldots,4$.  To reconstruct the Hamiltonian~(\ref{eq:tildeH}),
we must determine the coefficients $s_{k\ell;\nu}$ by solving~(\ref{eq:Ms}).

Each $M_{k\ell}$ is a real symmetric matrix whose off-diagonal elements
$M_{k\ell;\mu\nu}$, $\mu\neq\nu$, are determined by Eq.~(\ref{eq:Mkell}).  The
diagonal elements $M_{k\ell;\mu\mu}$ are unknown.  However, we know that
$M_{k\ell}$ should be a projector onto the 1D space spanned by
$\vec{s}_{k\ell}$, and the second equation in (\ref{eq:Ms}) determines the norm
of $\vec{s}_{k\ell}$ as well as the vector of diagonal elements
$(M_{k\ell;\mu\mu})_{\mu=0}^{\mu=3}$.  Thus, to determine the diagonal elements
of $M_{k\ell}$ and the corresponding eigenvector $\vec{s}_{k\ell}$, we note that
a rank-$1$ projector $\Pi$ with matrix entries $(g_{mn})$ must satisfy the
condition
\begin{equation}
  dg_{mn} \equiv  g_{mm} g_{nn} - g_{mn}^2 = 0 \quad \forall m,n.
\end{equation}
Thus, given the off-diagonal elements of $M_{k\ell}$, we choose the
diagonal elements of $M_{k\ell}$ such as to minimize the norm of the
error $e=\sum_{m,n} dg_{mn}^2$, and take $\vec{s}_{k\ell}$ to be the
eigenvector corresponding to the eigenvalue of $M_{k\ell}$ closest to
$1$, normalized to ensure $\norm{\vec{s}_{k\ell}}_2^2= c_{k\ell}$.  It
is important to carefully choose the parameters for the optimization
here to ensure we find the diagonal elements corresponding to the global
minimum.  Ideally, the residual error $e$ should be $10^{-10}$ or less.

We implemented and tested the algorithm for our 100 Hamiltonians.  We
were able to correctly identify the level structures for all but one
case: system 73, which has two nearly identical transition frequencies
with $\omega_2=1.8012$ and $\omega_3=1.8026$, for $N=1025$ data points
sampled at $N_e=125$, $250$, and $500$ experiment repetitions per data
point.  Even for this system, we were able to correctly identify the
level structure by doubling the number of data points $N$, with the
exception of $N_e=250$ where at least $N=4097$ data points were needed.
Of course, in practice more data points would be required for such a
system to be confident that the identification is correct, as explained
earlier.

\begin{table}
\begin{tabular}
{|l|cp{1.1em}p{1.1em}|cp{1.1em}p{1.1em}|cp{1.1em}p{1.1em}|cp{1.1em}p{1.1em}|}
\hline
$N \backslash N_e$ & 125 & & & 250 && & 500 && & 1000 && \\
& $\E(H)$ & $1\%$ & $5\%$ 
& $\E(H)$ & $1\%$ & $5\%$ 
& $\E(H)$ & $1\%$ & $5\%$ 
& $\E(H)$ & $1\%$ & $5\%$ \\\hline
16,385 & 0.40 & 11 & 1  & 0.27 & 5 & 0 & 0.18 &  2 & 0 & 0.13 &  4 & 0\\
 8,193 & 0.57 & 22 & 0  & 0.41 & 8 & 0 & 0.31 &  8 & 1 & 0.19 &  4 & 0\\
 4,097 & 0.87 & 41 & 5  & 0.66 & 25& 2 & 0.41 & 15 & 1 & 0.28 &  7 & 1\\
 2,049 & 1.12 & 60 & 7  & 0.91 & 45& 6 & 0.58 & 19 & 4 & 0.44 & 12 & 2\\
 1,025 & 1.81 & 81 & 13 & 1.32 & 64& 8 & 0.84 & 34 & 5 & 0.63 & 31 & 4\\\hline
\end{tabular}
\caption{Relative error $\E(H)=100\times\norm{H^{\est}-H}/\norm{H}$ of
  reconstructed Hamiltonian (with phase corrections) in \%.  Each table entry
  consists of three numbers: the median error (in \%) and the number of
  systems (of 100) with relative error exceeding 1\% and 5\%, respectively.}  
\label{table:err_H}
\end{table}

To gauge the overall accuracy of the estimated Hamiltonians we would
like to compute the norm of the error $\norm{\Delta H}=\norm{H^{\est}
-H^{\act}}$, or the relative error $\norm{\Delta H}/\norm{H^{\act}}$, 
where we choose the operator norm here.  However, calculating the norm 
of the error is complicated by the fact that we can only reconstruct 
the Hamiltonian up to the diagonal matrix $D$ and energy inversion 
symmetry.  Thus we must compensate for the phases that are ``unobservable'' 
in our model by setting
\begin{equation}
  \label{eq:DeltaH}
  \norm{\Delta H} = \norm{D^\dag H^{\est} D -H^{\act}}
\end{equation}
with $D=\diag(1,\delta_{12},\delta_{13},\delta_{14})$, where 
\begin{equation}
  \delta_{1\ell} = \angle(H^{\act}_{1\ell}) - \angle(H^{\est}_{1\ell}),
  \quad \ell=2,3,4,
\end{equation}
and $\angle(H_{1l}^{\act})$ is the complex phase of the $(1,l)$ matrix
element of $H^{\act}$, etc.  Table~\ref{table:err_H} shows the results
of the percentage relative errors $\norm{\Delta H}/\norm{H^{\act}}$
for our 100 test systems, for different values of $N$ and $N_e$.
Medians of the relative errors range from $0.13$\% for $N=16385$ and
$N_e=1000$ to $1.81$\% for $N=1,025$ and $N_e=125$.

\section{Control Hamiltonian Tomography}
\label{sec:5}

We have seen that our procedure can characterize a single Hamiltonian up
to a (physically irrelevant) global energy shift, and three relative
phases $\delta_{1n}$ for $n=2,3,4$, due to the freedom to redefine each
of the measurement basis vectors by a $\U(1)$ phase minus an overall
phase.  If we can only measure the system in a fixed basis and prepare
it in the measurement basis states, and the evolution is determined by a
single fixed Hamiltonian, then we have determined all observable
parameters.  However, for the system to be controllable, we require at
least two (noncommuting) Hamiltonians, or more generally we must have
the ability to modify the Hamiltonian by changing control parameters
$\vec{f}$, e.g., by applying external fields or varying applied gate
voltages, etc.  In this case we can still choose the phases
$\delta_{1n}^{(0)}$ for one ``reference'' Hamiltonian $H_0=H(\vec{f}_0)$
as we wish, e.g., $\delta_{1n}^{(0)}=0$ but the phases
$\delta_{1n}^{(\vec{f})}$ for all other Hamiltonians $H(\vec{f})$ are
now observable and thus relevant, and complete control Hamiltonian
reconstruction therefore requires that we identify them.

To achieve this, note that if can initialize the system in the
superposition state $\ket{\Phi}=\sum_{j=1}^4 \alpha_j\ket{j}$ and
measure the time-evolved state 
\begin{equation}
  \ket{\Phi(t)} = U_{\vec{f}}(t)\ket{\Phi}
 =D_{\vec{f}}^\dag \tilde{U}_\vec{f}(t)D_{\vec{f}}\ket{\Phi}
\end{equation}
with $U_{\vec{f}}(t)=\exp[-it H(\vec{f})]$,
$\tilde{U}_{\vec{f}}(t)=\exp[-it \tilde{H}(\vec{f})]$ then
\begin{equation}
 \label{eq:psuper}
 p_{\ell}(t)
  = |\bra{\ell}D_{\vec{f}}^\dag \tilde{U}_\vec{f}(t)] D_{\vec{f}}\ket{\Phi}|^2
  = |\bra{\ell}\tilde{U}_\vec{f}(t) D_{\vec{f}}\ket{\Phi}|^2
\end{equation}
shows that the phases $\delta_{1n}^{\vec{f}}$ that determine
$D_{\vec{f}}=\diag(1,e^{i\delta_{12}^{\vec{f}}},
e^{i\delta_{13}^{\vec{f}}},e^{i\delta_{14}^{\vec{f}}})$ are now
observable as $D_\vec{f}$ no longer commutes with the initial state
$\ket{\Phi}$.  As $\tilde{U}_\vec{f}(t)$ is fully determined by previous
steps, if the initial state $\ket{\Phi}$ is known, then the only unknown
parameters in Eq.~(\ref{eq:psuper}) are $\delta_{1n}^B$ for $n=2,3,4$.
Given a set of measured values $d_{\ell k}$ for $p_\ell(t_k)$, we can
determine the unknown parameters $\delta_{1n}^B$ by minimizing the
least-squares error
\begin{equation}
 \label{eq:err_delta}
  \vec{e} = \sum_{\ell=1}^4 \norm{\vec{p}_\ell-\vec{d}_\ell}_2^2.
\end{equation}
where $\vec{p}_\ell=(p_\ell(t_0),\ldots,p_\ell(t_K))$ and
$\vec{d}_\ell=(d_{\ell 0},\ldots,d_{\ell K})$ for $\ell=1,2,3,4$.  An
explicit expression for $p_\ell(t)$ derived in Appendix~\ref{app:B}
shows that we can in principle determine all the phases if the initial
state satisfies $\alpha_j\neq 0$ for all $j$~\footnote{In this case, the
small number of parameters to be estimated means that a simple
least-squares fit performs adequately. A Bayesian analysis is made more
difficult by the complicated dependence of the basis functions on the
undetermined parameters, though this approach may lead to better
estimates than the ones presented here, though at a cost of greater
computational complexity.  This is a topic for further research.}.
Moreover, it is advantageous to choose a balanced initial state,
$|\alpha_j|^2\approx \frac{1}{4}$ for all $j$, if possible, to maximize
signal to noise ratios.

\begin{figure}
\includegraphics[width=0.9\columnwidth]{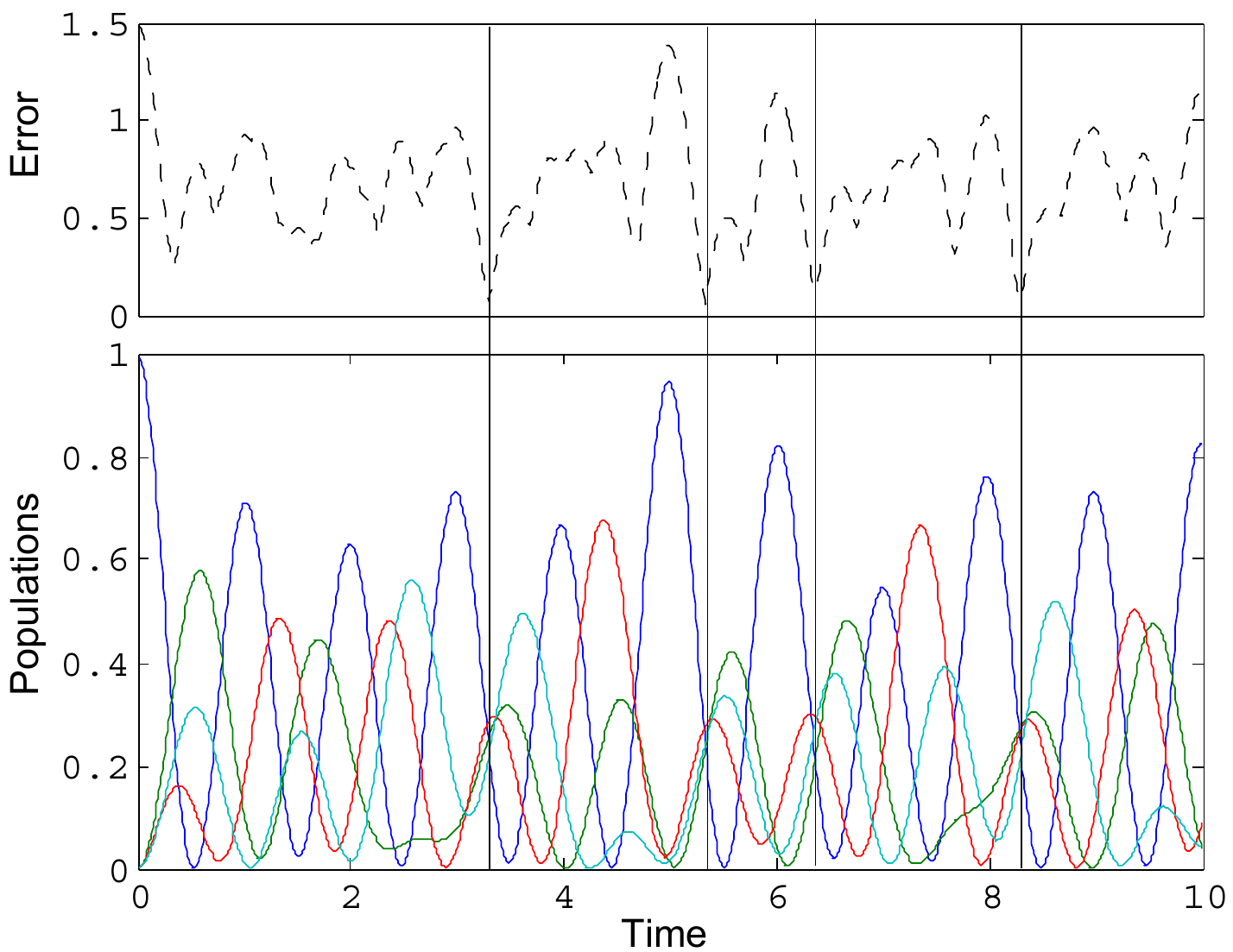} \caption{(Color
online) Evolution of populations $|\alpha_j(t)|$ under reference
Hamiltonian (system 5) and error $\sum_j
\left||\alpha_j|^2-\frac{1}{2}\right|$ from ideal balanced initial
state.  We selected the second minimum (which is the global minimum for
$0\le t\le 10$) at $t=5.34$ as initial evolution time $t_*$ for the
$\delta$ estimation step.}  \label{fig:Phi_t}
\end{figure}

\begin{figure}
  \includegraphics[width=0.9\columnwidth]{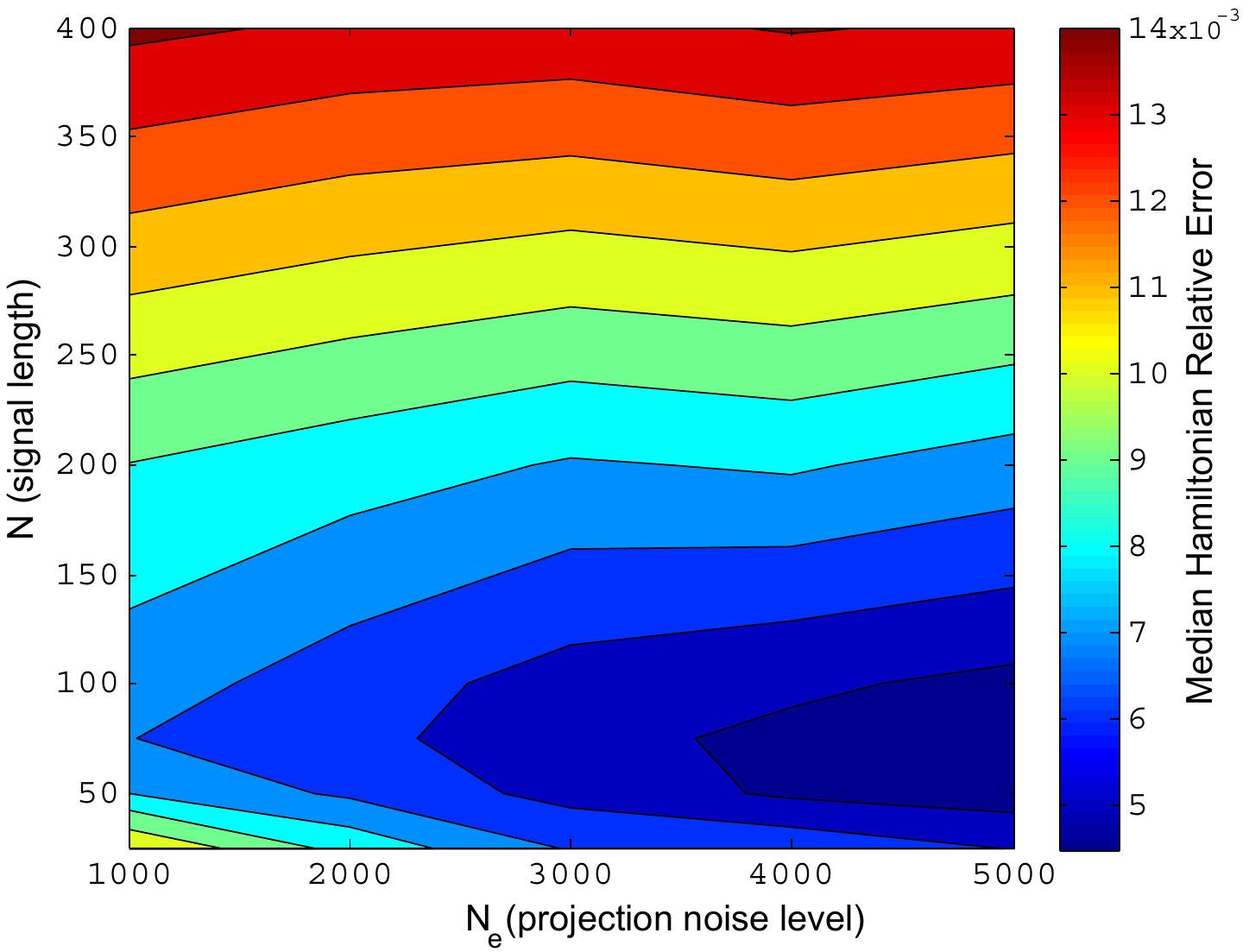} \caption{(Color
    online) Median relative error of estimated Hamiltonian after $\delta$
    estimation as a function of signal lengths $N$ and projection noise level
    $N_e$ of the measured traces $p_\ell(t_k)$.  In all cases the most accurate
    estimates for $\tilde{H}$ from the previous step, i.e. $N=16385$
    \& $N_e=1000$, were used.}  \label{fig:PhaseEstimContour}
\end{figure}

To prepare such an initial state, we can use the reference Hamiltonian
$H_0$.  Unless the reference Hamiltonian is such that one or more of the
measurement basis states are completely decoupled from state $\ket{1}$,
it is almost certain that the time-evolved state
$\ket{\Phi_1(t)}=U_0(t)\ket{1}=\sum_{j=1}^4\alpha_j(t)\ket{j}$ with
$U_0(t)=\exp(-it H_0)$ will satisfy $\alpha_j(t)\neq 0$ for all $j$ for
at least some $t>0$.  Thus, having characterized the Hamiltonians
$H=H(\vec{f})$ for different control settings $\vec{f}$ up to the phases
$\delta_{1n}^{\vec{f}}$, all we need to do is to select a suitable
reference Hamiltonian $H_0=H(\vec{f}_0)$, and find a time $t_*$ such
that the time-evolved state $\ket{\Phi_1(t_*)}$ satisfies
$|\alpha_j(t_*)|\approx \frac{1}{2}$.  This is generally not difficult.
For instance, we randomly choose the Hamiltonian for test system $5$ as
our reference Hamiltonian.  Fig.~\ref{fig:Phi_t} shows that there are
several times $t\in[0,10]$ at which the populations $|\alpha_j(t)|^2$ of
all levels (in the measurement basis) are approximately equal.  We pick
one of these times $t_*=5.34$, set $\ket{\Phi}=U_0(t_0)\ket{1}$, and
obtain the measurement traces $p_\ell(t_k)$ as follows:
\begin{enumerate}
\item Initialize system in measurement basis state $\ket{1}$.
\item Let it evolve under Hamiltonian $H_0$ for time $t_*$.
\item Change control settings to $\vec{f}$ and let system evolve
      for $t$ time units under Hamiltonian $H_{\vec{f}}$.
\item Perform measurement $\Rightarrow$ outcome $\ell=1,2,3,4$.
\end{enumerate}
As before we repeat this experiment $N_e$ times for a fixed $t$ to
estimate $p_\ell(t_k)$ (number of times the outcome was $\ell$ divided
by $N_e$), and then repeat for different times $t_k$ to obtain estimates
for $p_\ell(t_k)$.

\begin{table}
\begin{tabular}
{|l|cp{1.1em}p{1.1em}|cp{1.1em}p{1.1em}|cp{1.1em}p{1.1em}|cp{1.1em}p{1.1em}|}
\hline
$N \backslash N_e$ & 125 & & & 250 && & 500 && & 1000 && \\
& $\E(H)$ & $1\%$ & $5\%$ 
& $\E(H)$ & $1\%$ & $5\%$ 
& $\E(H)$ & $1\%$ & $5\%$ 
& $\E(H)$ & $1\%$ & $5\%$ \\\hline
16,385 
        & 0.89 & 40 &  2 & 0.70 & 27 &  1 & 0.56 & 14 &  1 & 0.45 &  6 & 1\\
        & 2.12 & 78 & 12 & 1.54 & 75 &  9 & 1.00 & 50 &  6 & 0.76 & 35 & 1\\\hline
 8,193 
        & 1.09 & 54 &  5 & 0.84 & 35 &  0 & 0.69 & 28 &  1 & 0.53 & 17 & 1\\
        & 2.60 & 89 & 24 & 2.21 & 86 & 14 & 1.66 & 77 & 10 & 1.08 & 52 & 5\\\hline
 4,097 
        & 1.48 & 68 &  7 & 1.12 & 58 &  5 & 0.91 & 43 &  5 & 0.61 & 26 & 2\\
        & 3.47 & 96 &  41& 3.13 & 94 & 27 & 2.20 & 87 & 13 & 1.45 & 65 & 7\\\hline
 2,049 
        & 2.24 & 88 & 15 & 1.45 & 74 &  7 & 1.12 & 55 &  7 & 0.78 & 37 & 4\\
        & 6.06 & 98 & 67 & 3.91 & 93 & 37 & 2.85 & 89 & 25 & 2.25 & 78 & 16\\\hline
 1,025 
        & 3.14 & 95 & 29 & 2.44 & 90 & 18 & 1.60 & 80 &  8 & 1.22 & 59 & 6\\
        & 8.36 & 98 & 76 & 5.92 & 96 & 59 & 4.50 & 95 &  48& 3.00 & 88 & 32\\
\hline
\end{tabular}
\caption{Relative error $\E(H)=100\times\norm{H^{\est}-H}/\norm{H}$ of
  reconstructed Hamiltonian with estimated phases $\delta_{1n}$ in \%
  (no phase corrections).  As before, each table entry consists of three
  numbers: the median error (in \%) and the number of systems (of 100)
  with relative error exceeding 1\% and 5\%, respectively.  The first
  row in each box are the estimates obtained for signals $p_\ell(t_k)$ of
  length $N=50$, the second row the estimates obtained for signals of
  length $N=200$, in both cases sampled at $N_e=5000$.}
  \label{table:err_H2}
\end{table}

We tested the phase estimation procedure for the estimated Hamiltonians
obtained in the previous step.  For each of the $100$ systems we first
generated (simulated) measurement signals for $p_\ell(t_k)$ of varying
length $T=(N-1)\Delta t$ and levels of projection noise $N_e$.  The
number of points ranged from $N-1=25$ to $1000$ data points, sampled at
$\Delta t=0.1$ fixed as before; the measurement repetitions $N_e$ from
$1000$ to $5000$. In the reconstruction of the phases, we only assume we
know the estimated $H_0$, hence the estimated $\ket{\Phi(0)}$, and the
estimated $H_\vec{f}$, determined in Section~\ref{sec:4}. While the most
accurate estimates for the frequency and linear coefficient estimation
step (step 1) were obtained for the longest signals ($N=16,385$), we
find that the accuracy of the phase estimation step peaks at around
$N\approx 50$, and that longer signals are in fact highly detrimental
(Fig.~\ref{fig:PhaseEstimContour}).  This may seem very surprising at
first but can be at least partly explained by the fact that even small
inaccuracies in the initial estimates, especially for the frequencies,
will accumulate over time and increase the discrepancy between the
projected evolution of the system based on our Hamiltonian estimates and
the true evolution.

Based on these results we settled for signals of length $N-1=50$ with
$N_e=5000$ measurement repetitions per data point for the final phase
estimation step.  For each of the $2000$ estimated Hamiltonians
$\tilde{H}$ obtained in the first step --- corresponding to the $100$
different test systems, as well as four levels of projection noise
$N_e\in\{125,250,500,1000\}$ and five signal length $T=(N-1)\Delta t$
for $N-1\in\{2^{10},2^{11},2^{12},2^{13},2^{14}\}$ with $\Delta t=0.1$
fixed, each --- we estimated the phases $\delta_{1n}$, and used the
results to reconstruct the total Hamiltonian $H=D^\dag \tilde{H} D$.
Table~\ref{table:err_H2} shows the results in terms of the median of
relative errors.  For comparison we include the Table the results
obtained had signals of length $N=200$ been used instead.  Comparison of
the numbers clearly shows that longer signals are detrimental for the
phase estimation step.  In addition to substantially decreased accuracy,
longer signals also slowed down the numerical optimization, making it
more difficult for the routine to find the global minimum.  In view of
the complicated dependence of $p_\ell(t_k)$ (see appendix \ref{app:B})
on the parameters $\delta_{1n}$, $n=2,3,4$, we initially explored
population-based (global) optimization strategies, especially
evolutionary algorithms, but found that it was substantially slower and
far \emph{less} effective in finding the global minimum of the error
Eq.~(\ref{eq:err_delta}) than a gradient-based (BFGS-type) local
optimization algorithm.  In fact, for short signals the local
optimization routine generally succeeded in finding the global minimum
in a single run, starting with a random guess for
$\gvec{\delta}=(\delta_{12},\delta_{13},\delta_{14})$, although the
optimization was repeated with several different initial guesses to
increase the probability that we had indeed found the (globally) best
value for $\gvec{\delta}$.

\section{Concluding Discussion}
\label{sec:6}

We have presented a method for characterizing the Hamiltonian and its
dependence on external control parameters, which is a pre-requisite for
Hamiltonian Engineering and coherent control of the system's evolution,
for a generic two-qubit system, assuming only the ability of preparation
and measurement in a fixed basis.  Analysis of simulated measurement
data shows that the task of estimating the parameters from the complex,
noisy measurement signals with multiple frequencies, and reconstructing
the Hamiltonian is very challenging, and requires a carefully designed
multi-step approach, combining spectral analysis, Bayesian analysis and
several carefully designed optimization steps to reconstruct the energy
level structure and matrix representation of the Hamiltonian.  In the
absence of any control, the Hamiltonian can only be reconstructed up to
three phases, due to the freedom to redefine the measurement basis by
$\U(1)$ phase rotations.  This symmetry can be broken if the system can
be prepared initially in a suitable superposition state, and we exploit
this fact to achive full control Hamiltonian tomography in a simple two
step procedure.

The Bayesian analysis assumes a Gaussian noise profile which, though not
strictly accurate, works well, especially in the large $N_e$ limit.  Any
significant deviations from Gaussian noise (e.g. Poissonian statistics
for small $N_e$ for $p\approx 0,1$) will tend to make the log-likelihood
estimates worse, and thus our estimates of the confidence that the model
fits the data are conservative~\cite{88Bretthorst}.  More accurate error
estimates could be obtained using Bayesian analysis with a Poissonian
noise model, though our results show that even a Gaussian noise model
results in a huge improvement of two orders of magnitude or more in the
accuracy of the frequency estimates, compared to estimates obtained from
simple spectral analysis.  This turned out to be crucial for successful
Hamiltonian reconstruction.  The frequency estimates obtained from the
power spectrum combined with a simple least-squares error minimization
to find the optimal spectral amplitudes proved to be too inaccurate for
Hamiltonian reconstruction, leading to inconsistent equation systems and
significant errors, and any attempt to obtain estimates of the
parameters by direct minimization of the least-squares error of the
measurement signals and the expected signals resulted in reconstructed
Hamiltonians that were little better than random for our test systems.

Though we have implicitly assumed a Hamiltonian model, i.e., that
incoherent effects will be negligible on the time scales of interest,
any significant deviation from the assumed model, e.g., significant
decoherence or coupling to additional states outside the two-qubit
subspace would result in low likelihoods of the chosen (four-level)
Hamiltonian model.  Such effects can easily be incorporated into the
analysis by changing the basis functions, e.g., using damped
exponentials instead of sinusoids or including additional states, which
we will consider in further work.  Furthermore, any prior information
about the structure of the Hamiltonian can be incorporated to make the
Bayesian analysis more efficient.  Thus, the method lends itself to
adaptive protocols, as we can adaptively sample the system until certain
targets for the likelihood or error estimates are met, ensuring that we
perform enough measurements to get accurate estimates but no more than
necessary.~\footnote{Composite pulses or robust optimal control can be
used to relax the need for extremely accurate Hamiltonian
characterization.}  This is especially important as the number of
measurements required will vary depending on the system.  For instance,
for a system with well spaced transition frequencies, a sharply peaked
likelihood function with a clearly identifiable global maximum can be
obtained with much less data than for a system with two almost
degenerate transition frequencies.

For control Hamiltonian tomography, the small but non-zero inaccuracies
in the initial estimation step lead to an optimum sampling time for the
second step due to divergence of the model from the true system behavior
at longer times.  In principle, it should be possible to use this
divergence to improve the initial estimates of the Hamiltonians, and
exploring such refinements could be an interesting avenue for future
research.  Errors in the second step decreased with increased signal to
noise ratio (increasing $N_e$), as the estimate of the phase parameters
does not depend on the signal length, unlike frequency resolution.  It
would also be interesting to investigate the accumulation of errors in
this multi-step estimation, especially how uncertainties in prior steps
affect the accuracy of the Bayesian estimation in subsequent stages.
Finally, in this paper we have dealt with the generic case.  When the
Hamiltonian has exact degeneracies then the measurement signals will
contain fewer than six frequencies.  In this case, the level structure
reconstruction becomes harder as the number of special sub-cases
increases and we may not be able to uniquely identify the Hamiltonian.
Although the set of Hamiltonians with exact degeneracies is of measure
zero, further study of these special cases may be of interest as one may
want to specifically engineer Hamiltonians with such level structures.

\begin{acknowledgments}
  We thank S. J. Devitt and J. H. Cole for discussions. SGS acknowledges
  funding from EPSRC Advanced Research Program Grant RG44815, EPSRC QIP
  Interdisciplinary Research Collaboration (IRC) and Hitachi.  DKLO is
  supported by the Scottish Universities Physics Alliance (SUPA) and the
  Quantum Information Scotland network (QUISCO). This work is supported
  by the National Research Foundation \& Ministry of Education,
  Singapore.
\end{acknowledgments}

\begin{appendix}

\section{Measured probabilities}
\label{app:B}

If the system is initialized in the generic superposition state
$\ket{\Phi}=\sum_{j=1}^4 \alpha_j\ket{j}$ and measured after evolving
for $t$ time units under the Hamiltonian $H_B=D^\dag\tilde{H}_B D$, then
the general expression for the probability $p_\ell(t)$ of measurement
outcome $\ell$ is
\begin{widetext}
\begin{align}
 p_\ell(t)
 &= |\bra{\ell}\sum_{\mu=1}^4e^{-i\lambda_{\mu}t}
    \ket{\xi_\mu}\bra{\xi_\mu}D\ket{\Phi}|^2 \nonumber\\
 &= \sum_{\mu,\nu=1}^4 \sum_{m,n=1}^4 \alpha_m\alpha_n^{*}
    e^{-i(\omega_{\mu\nu}t-(\delta_{1m}^B-\delta_{1n}^B))}
    \ip{\ell}{\xi_\mu}\ip{\xi_\mu}{m}\ip{n}{\xi_\nu}\ip{\xi_\nu}{\ell}
    \nonumber\\
 &= \sum_{\mu,\nu=1}^4\sum_{m,n=1}^4 |\alpha_m||\alpha_n|s_{\ell m;\mu}s_{n\ell;\nu}
    e^{-i(\omega_{\mu\nu}t-(\delta_{1m}^B-\delta_{1n}^B)-
    (\phi_m-\phi_n)-\delta_{\ell m;\mu}-\delta_{n\ell;\nu})} \nonumber\\
 &= 
   \sum_{\mu=1}^4 \Bigg[\sum_{m=1}^4 |\alpha_m|^2 s_{\ell m;\mu}^2
  +\sum_{m>n} 2|\alpha_m| |\alpha_n|s_{\ell m;\mu}s_{n\ell;\mu}
              \cos((\delta_{1m}^B-\delta_{1n}^B)+(\phi_m-\phi_n)+\delta_{\ell m;\mu}
  +\delta_{n\ell;\mu}))\Bigg]
\nonumber\\
 & +\sum_{\mu>\nu}\Bigg[ \sum_{m=1}^4 2|\alpha_m|^2 s_{\ell m;\mu}s_{m\ell;\nu}
   \cos(\omega_{\mu\nu}t-\delta_{\ell m;\mu}-\delta_{m\ell;\nu})\nonumber\\
 & \qquad+ 
   \sum_{m\ne n} 2|\alpha_m||\alpha_n|s_{\ell m;\mu}s_{n\ell;\nu}
   \cos(\omega_{\mu\nu}t-(\delta_{1m}^B-\delta_{1n}^B)-(\phi_m-\phi_n)-
        \delta_{\ell m;\mu}-\delta_{n\ell;\nu})\Bigg].
\end{align}
\end{widetext}
\end{appendix}

\end{document}